\documentclass[useAMS,usenatbib]{mn2e}

\usepackage{natbib}
\usepackage{amssymb,amsmath,upgreek}
\usepackage{rotfloat}
\usepackage{caption}

\usepackage{epsfig,graphicx,times,subfigure}
\usepackage{rotating}

\newcommand{\kms}{\mbox{km s$^{-1}$}}

\newcommand{\galmass}{$\mathrm{log}(\mathrm{M}_{\mathrm{host}}/\mathrm{M}_{\odot})$}
\newcommand{\magpergalmass}{$\mathrm{mag}/\mathrm{log}(\mathrm{M}_{\mathrm{host}}/\mathrm{M}_{\odot})$}

\newcommand{\aaps}{A\&A}
\newcommand{\aj}{AJ}
\newcommand{\pasp}{PASP}
\newcommand{\prd}{Phys.Rev.D}
\newcommand{\apj}{ApJ}
\newcommand{\apjl}{ApJL}
\newcommand{\mnras}{MNRAS}
\newcommand{\araa}{ARA\&A}
\newcommand{\apjs}{ApJS}

\newcommand{\nofinal}{581}

\title[SN Ia hosts]{How SN Ia host-galaxy properties affect cosmological parameters}
\author[Campbell, Fraser \& Gilmore]{H. Campbell$^{1}$\thanks{E-mail:hcc@ast.cam.ac.uk, mf@ast.cam.ac.uk},
	M. Fraser$^{1}$,
	G. Gilmore$^{1}$\\
$^{1}$Institute of Astronomy, University of Cambridge, Madingley Road, Cambridge, CB3 0HA, UK\\
}

\begin{document}

\date{Accepted to Monthly Notices of the Royal Astronomical Society}

%\pagerange{\pageref{firstpage}--\pageref{lastpage}} \pubyear{}

\maketitle

%\label{firstpage}

\begin{abstract}
We present a systematic study of the relationship between Type Ia Supernova (SN Ia) properties, and the characteristics of their host galaxies, using a sample of 581 SNe~Ia from the full Sloan Digital Sky Survey II (SDSS-II) SN Survey. We also investigate the effects of this on the cosmological constraints derived from SNe~Ia. Compared to previous studies, our sample is larger by a factor of $>4$, and covers a substantially larger redshift range (up to $z\sim0.5$), which is directly applicable to the volume of cosmological interest. We measure a significant correlation ($>5\sigma$) between the host-galaxy stellar-mass and the SN~Ia Hubble Residuals (HR). We find a weak correlation ($1.4\sigma$) between the host-galaxy metallicity as measured from emission lines in the spectra, and the SN~Ia HR. We also find evidence that the slope of the correlation between host-galaxy mass and HR is $-0.11$ \magpergalmass\ steeper in lower metallicity galaxies. We test the effects on a cosmological analysis using both the derived best-fitting correlations between host parameters and HR, and by allowing an additional free parameter in the fit to account for host properties which we then marginalize over when determining cosmological parameters. We see a shift towards more negative values of the equation of state parameter $w$, along with a shift to lower values of $\Omega_\mathrm{m}$ after applying mass or metallicity corrections. The shift in cosmological parameters with host-galaxy stellar-mass correction is consistent with previous studies. We find a best-fitting cosmology of $\Omega_{\mathrm{m}} =0.266_{-0.016}^{+0.016}$,  $\Omega_{\Lambda}=0.740_{-0.018}^{+0.018}$ and $w=-1.151_{-0.121}^{+0.123}$ (statistical errors only).

\end{abstract}

\begin{keywords}
   supernovae: general, cosmological parameters, dark energy, cosmology: distance scale,
\end{keywords}

\section{Introduction}

Type Ia Supernovae (SNe~Ia) arise from the explosion of a degenerate carbon-oxygen white dwarf, following either accretion from a non-degenerate companion, or a merger with a white dwarf secondary in a binary system \citep{Hil00}. Their relatively uniform peak luminosities has led to their use as standardisable candles for cosmology, where they can be used to probe the expansion of the Universe, and the acceleration of this expansion due to dark energy \citep[e.g.][]{Rie98,Sch98,Per99,Sul11}.

The absolute magnitude of an SN~Ia at maximum light correlates strongly with the rate of decline seen in the {\it B}-band after peak \citep[the light-curve `stretch', ][]{Phi93}, and with SN~Ia colour \citep{Rie96}. By applying empirical calibrations to a large sample of SNe~Ia, the intrinsic dispersion in their peak magnitudes is sufficiently reduced that they can be used to accurately derive cosmological parameters. More recently, the availability of precise, well-calibrated photometry for large samples of SNe Ia has motivated searches for additional correlations between Type Ia SNe and their spectroscopic properties \citep[e.g.][]{Fol11} or host-galaxy characteristics \citep[e.g.][]{Sul10,Chi13a}. The identification of any such correlations can be used to further reduce the scatter in the Hubble diagram, and improve estimates of the Hubble constant H$_0$ and equation of state $w$. Identifying relationships between SNe~Ia properties and their host galaxies can also help shed light on the progenitor systems and physical mechanisms which lead to SNe~Ia \citep[e.g.][]{Mag13}.

Previous SN studies, such as \citet{Kel10}, \citet{Sul10}, \citet{Lam10a}, \citet{DAn11}, \citet{Li11}, \citet{Gup11}, \citet{Joh13}, \citet{Chi13a}, \citet{Pan14} have shown that there are correlations between the peak brightness of an SN~Ia, and certain properties of its host galaxy. Of those, the correlation between host-galaxy stellar mass and SNe~Ia brightness (after correction for SN stretch and colour) has been investigated the most. \citet{Kel10} have shown that more massive galaxies tend to host SNe~Ia that are $\sim$10 per cent brighter after light-curve corrections at 2.5$\sigma$ confidence. \citet{Sul10} demonstrated that separating a sample of SNe~Ia according to whether they had a low or high-mass host galaxy, and using two different values of $M$ (the peak absolute magnitude in the distance modulus calculation) for these samples improves the precision of the fitted cosmological parameters by $3.8-4.5\sigma$. They found that the absolute value of the offset is 0.08 mag at 10$^9$ M$_\odot$, with 4$\sigma$ confidence.

It is quite likely, however, that the host-galaxy mass is merely a proxy for an underlying physical property such as metallicity, as any individual SN~Ia should be `unaware' of (and hence unaffected by) the total mass of its host. Several previous studies have investigated the host-galaxy metallicity, including \citet{DAn11}, \citet{Joh13}, \citet{Chi13a}, \citet{Pan14}, who all found that SNe~Ia in higher metallicity galaxies are over luminous for their light-curve shape and that their Hubble Residual (HR; the difference between the measured distance modulus and that expected from the best-fitting cosmology) are $\sim$0.1 mag brighter, at confidence levels varying between $<2.5\sigma$ \citep{Joh13}, $2.5\sigma$ \citep{Pan14}, $2.9\sigma$  \citep{Chi13a} and $>4\sigma$ \citep{DAn11}. However, as spectroscopy is required to measure galaxy metallicity, it is a much harder property to measure than mass, and hence samples are smaller.

Other studies have investigated regions of local star formation. \citet{Rig13} used the SN Factory sample, while \citet{Rig15} used the Constitution sample to investigate areas of local star formation using Galaxy Evolution Explorer (GALEX) FUV/NUV data. They showed that SNe in locally star-forming environments are on average $0.094\pm0.037$ mag fainter than SNe~Ia having locally passive environments. They also caution that if the ratio of SNe~Ia in local star-forming environments changes with redshift or sample selection, this can lead to a bias in cosmological measurements. However, \citet{Kel15} show that the distances to SNe in locally star formation regions can be calibrated to $<4$ per cent. They suggest that the smaller scatter in this sample is due to only one progenitor type erupting in these regions. However, \citet{Jon15} see no correlation between the  regions of local star formation and the SN parameters.

We also note that some authors have found the correlation between host-galaxy mass and HR to be much less significant than that found by \cite{Sul10} and others. Rather than matching to a template, \cite{Kim14} fit SNe~Ia light curves by modelling them as stochastic functions described by Gaussian Processes. Using this different technique for fitting SNe~Ia, they find no evidence for host-galaxy mass to HR relation. The residual step at 10$^{10}$  M$_\odot$ is 0.013$\pm$0.031 mag, which is consistent with zero. They interpret the absence of a correlation as a result of their technique of light curve fitting, which they argue can better account for diversity in SNe~Ia.

In this paper, we use the photometrically-classified sample of SNe~Ia from the Sloan Digital Sky Survey II (SDSS-II) SN Survey, presented by \cite{Cam13}, to investigate correlations between the properties of the host galaxies of SNe~Ia, and the properties of the SNe~Ia themselves. In Sections \ref{sect:data} and \ref{sect:prop_dist}, we introduce the data and techniques used; in Section \ref{sect:results}, we present the analysis of possible correlations; in Section \ref{sect:subsample}, we discuss analyses of subsets of the data, while in Sections \ref{sect:cosmo} and \ref{sect:discussion}, we calculate the effect of these correlations on derived cosmological parameters, and discuss their implications.

We note that a paper by \cite{Wol15} has recently been submitted, which seeks to address some of these same questions using the SDSS-II data set. However, there are significant differences between these two papers; \citeauthor{Wol15} focus on performing a careful reanalysis of all the host-galaxy properties (such as metallicity and mass), while we use the standard SDSS products. In this work, we also examine the effect of our results on cosmological fits. Two independent analyses of the same data also function as a useful check on the reliability of the results obtained; we discuss this further in Section~\ref{sect:wolf}.

\section{Data and methodology}
\label{sect:data}

\subsection{SDSS-II SN sample}

The SDSS-II SN Survey \citep{Fri08,Sak08,Sak14} was a dedicated search for intermediate-redshift SNe~Ia between 2005 and 2007, in a 300 deg2 field called `Stripe 82'. The survey was carried out in multicolour ({\it ugriz}) imaging, for three months per year, on the SDSS 2.5m telescope \citep{Gun98}. After three years of observations, more than 500 SNe~Ia had spectroscopic confirmation \citep{Zhe08,Kon11,Ost11}. The spectroscopically confirmed sample of SDSS-II SNe~Ia has now been used to constrain cosmological parameters both independently \citep{Kes09, Sol09,Lam10a} and in a joint analysis with the Supernova Legacy Survey \citep[SNLS;][]{Bet14}. The SDSS-II SN sample has also been used to measure the SN~Ia rate \citep{Dil08, Dil10, Smi12}, examine the rise-time distribution \citep{Hay10} and study the correlations between SNe~Ia and their host galaxies \citep{Lam10b, DAn11, Gup11, Gal12, Hay13} and spectroscopic indicators \citep{Kon11, Nor11, Fol12}.

SDSS also identified a large sample of potential SNe~Ia which were {\it not} spectroscopically confirmed. \cite{Cam13} demonstrated that if these candidate SNe could be photometrically classified with sufficient efficiency and purity, then they could also be used for cosmological purposes. To make robust photometric classifications it is necessary to know the host-galaxy redshift. Therefore, an ancillary program was run as part of the SDSS-III Baryon Oscillation Spectroscopic Survey \citep[BOSS;][]{Daw13,Olm14} between 2009 and 2010 to obtain the spectra and redshifts of the host galaxies of a large sample of SN candidates detected by the SDSS-II SN Survey. Details of the target selection and data reduction for this sample of galaxies are outlined in \citet{Cam13}, and details of the data analysis and redshifts for the sample are presented in \citet{Olm14}. In total, 3520 redshifts were measured for the host galaxies of SN candidates (and other transients), to a limiting galaxy magnitude of ${\it r} < 22.0$ mag. 
 
A sample of 752 high-quality photometrically-classified SNe~Ia for use in cosmological analyses was constructed in \citet{Cam13}. This sample was selected on the basis of a Bayesian light curve classifier, {\sc PSNID} \citep{Sak11}, which uses SNe templates and fits to {\sc SALT2} templates \citep{Guy07}, combined with stringent data-quality cuts. The sample covers the redshift range of $0.05 < z < 0.55$. Using detailed survey specific simulations \citet{Cam13} estimate the completeness to be 70 per cent and the remaining contamination from non-Ia SNe to be $< 4$~per cent. This photometrically-classified SNe~Ia sample was shown to produce comparable and competitive constraints when compared to cosmological analyses from the SNLS spectroscopically-confirmed SNe~Ia sample \citep{Guy10, Sul10}.

We use the sample from \citet{Cam13} throughout this paper and the host-galaxy properties determined from the BOSS spectra. This sample is slightly different from the photometrically-classified sample presented in \citet{Sak14}, due to differences in host galaxy association and selection cuts in constructing the sample. However, to carry out cosmological analysis of such a magnitude limited sample it is necessary to correct for Malmquist bias, hence we use the \citet{Cam13} data set and the  Malmquist bias correction derived within. We also note that within the 752 photometrically-classified SNe~Ia in the  \citet{Cam13} sample, a subset of 208 SNe~Ia have an additional spectroscopic classification.

All the SNe~Ia have been fitted by the SALT2 light-curve fitter, this models the spectral energy distribution (SED) evolution of SNe~Ia and their intrinsic variation using SNe~Ia templates, and  parameterizes the SNe~Ia by three parameters; the stretch ($x_1$), colour and apparent magnitude. The stretch, $x_1$, is a fitting parameter which is used to scale the observed light curve of an SN~Ia to a template. The colour is defined by $c=(B-V)_\mathrm{max} - \langle B-V \rangle$.

The SALT2 parameters  from the light curve fits are used to calculate the distance modulus to each SN~Ia: 

\begin{equation}
\mu=m_\mathrm{B}-M+\alpha \times x_1-\beta \times c -\mu_\mathrm{corr}
\end{equation}

Where $\mu_\mathrm{corr}$ is the Malmquist bias correction, (in this paper, we show results with and without this correction, denoted HR$_\mathrm{corr}$ and HR$_\mathrm{uncorr}$, respectively) and which is defined by an analytic prescription laid out in \citet{Cam13} from modelling using SNe~Ia simulations with the SuperNova ANAlysis (SNANA) code \citet{Kes09}. 
The Malmquist bias correction is defined as:

\begin{equation}
\mu_\mathrm{corr}=ae^{(bz)}+c
\end{equation}
where $a = -0.004\pm0.001$, $b = 7.26\pm0.31$, and $c = 0.004\pm0.006$.

The parameters $\alpha, \beta$ and $M$ (absolute $B$-band magnitude at peak) are constants that can either be derived for the whole sample simultaneously with the best-fitting cosmology, or can be constrained from other data. In our cosmology analysis, presented in Section~\ref{sect:cosmo}, we follow the same procedure as in \citet{Cam13} and we allow $\alpha$ and $\beta$ to float within priors and analytically marginalize over $M$ (which is degenerate with $H_0$).

The HR for each SN is calculated by subtracting the best-fitting cosmology found in \citet{Cam13}  ($w=-0.96^{+0.1}_{-0.1}, \Omega_\mathrm{m} = 0.29^{+0.02}_{-0.02}, \Omega_{\Lambda}= 0.71^{+0.02}_{-0.02}$).

\subsection{Host-galaxy properties}
\label{host_properties}

We adopt the SDSS DR10 \citep{Ahn12} host-galaxy parameters for all the SNe~Ia hosts where these are available. Unfortunately, 150 of our SNe~Ia host galaxies are missing processed host-galaxy spectral properties, as these were taken during commissioning of BOSS and were not fully processed by the SDSS-II BOSS pipeline and thus are not in the DR10 Portsmouth `Stellar Kinematics and Emission Line Fluxes' tables \citep{Tho13} used in this analysis. As these 150 SNe are a random subset of the full distribution, they should not bias the results presented in this paper. 

Initially, host galaxies were divided into star-forming and passive categories, according to flags in the Portsmouth `Stellar Kinematics and Emission Line Fluxes' DR10 tables. These classifications (and further sub-classifications) are based on the galaxies location on a `Baldwin, Phillips \& Terlevich' (BPT) diagram. This is a plot of emission line flux ratios, ([OIII]$\lambda$5007)/(H$\beta \lambda$4861) against ([NII $\lambda$6583)/(H$\alpha \lambda$6563); \citealp{Bal81}). When an active galactic nucleus (AGN) is present in a galaxy, its line emission can dominate over the line flux from star formation, rendering measurements of the latter difficult. As a first attempt at removing AGNs, we applied the same thresholds as in \citet{Kew01}, which removed 154 host galaxies from our sample. 

%Galaxies which lie below the dividing line on the BPT diagram are undergoing some star-formation, while galaxies which lie above it most likely contain an active galactic nucleus (AGN)

We also investigated an alternative way of removing AGN from our sample, as many of our galaxy spectra have low S/N spectra and it is difficult to get significant detections of all four lines required to place a galaxy on a BPT diagram. \citet{Car01} suggested removing AGN using a `two-line' diagnostic. They define galaxies as having an AGN if log(([NII $\lambda$6583)/(H$\alpha \lambda$6563))$>-0.2$. \citet{Mil03} showed that all four lines are required to robustly classify star-forming galaxies but that the `two-line' diagnostic is adequate for removing AGN. This  `two-line' diagnostic only removed 20 galaxies as likely AGN hosts. We carried out the subsequent analyses with both AGN diagnostics and found consistent correlations between the host galaxy and SNe~Ia parameters for both samples. In the following sections, we show the results from the `two-line' AGN diagnostics. It is worth noting that many correlations became slightly stronger, as might be expected from larger sample sizes, however the metallicity correlation with the SNe~Ia HR becomes slightly less significant (from 1.8$\sigma$ to 1.4$\sigma$) possibly due to the inclusion of lower S/N data. After removing AGNs from our sample, using the `two-line' diagnostic, we are left with 543 star forming and 38 passive host galaxies. 

The stellar mass of a galaxy can be derived by comparing the observed broad-band photometry to the best-fit spectral energy distribution (SED) template. The grid of templates are based on stellar population models, and cover a range of physical parameters (i.e., ages, dust content, chemical composition). The stellar mass of all BOSS galaxies, including our SNe~Ia host galaxies, have been calculated using the \citet{Mar06} population synthesis models  \citep{Mar13}. The `HyperZspec' code used by \citet{Mar13} to calculate the masses is a modified version of `HyperZ' \citep{Bol00}, with the SED fitting performed at a fixed redshift, which is derived from the spectra. HyperZspec computes the $\chi^2_\mathrm{red}$ for a large number of templates, with varying star-formation histories, and identifies the best-fitting template.

There are four masses computed for each galaxy in SDSS DR10, one with the best-fitting passive model and one with the best-fitting star-forming model, for both a Salpeter and Kroupa Initial Mass Function (IMF). We use the fits from the Kroupa IMF throughout this paper. We use the stellar mass from the star-forming model for the host galaxies which have been classified as `star forming' or `composite'  and the passive model for all other host galaxies. 

We use the ages presented in \citet{Mar13} and again use the star-forming ages for the hosts with BPT flag `star forming' or `composite', and passive ages for others.

Measuring gas-phase metallicities from intermediate resolution and low S/N spectra is a long-standing problem. Ideally, the metallicity would be measured via the so-called `direct method', where the auroral [O~{\sc iii}] $\lambda$4363 line is used to determine the electron temperature of the emitting region, and forbidden emission lines are used to measure abundances. However, the [O~{\sc iii}] $\lambda$4363 line is typically only detectable for $Z<$~0.5~Z$_{\odot}$, as above this threshold the gas is cooled via metal lines in the IR, and the auroral lines cannot be measured. Instead, we have used the strong-line diagnostic O3N2 \citep{Pet04} to determine metallicities for our sample. O3N2 = log[([O~{\sc iii}]$\lambda$5007/H$\beta$) / ([N~{\sc ii}]$\lambda$6583/H$\alpha$)], and is valid over the metallicity range 8.1~$<$~12+log[O/H]~$<$~9.1 dex. The O3N2 diagnostic has several advantages, namely that it is single-valued\footnote{Some other diagnostics, such as the R23 method \citep{Pil01, Pil05, Lia07, Yin07}, are double-valued, with a degenerate high (12 + log[O/H] $>$ 8.5) and a low (12 + log[O/H] $<$ 8.5) metallicity solution for a given line ratio.}, and that it is largely unaffected by reddening, as the two line ratios [O~{\sc iii}]/H$\beta$ and [N~{\sc ii}]/H$\alpha$ rely on lines which are close in wavelength. 

For the emission line fluxes we again use the results from the SDSS DR10 \citep[the `Portsmouth results';][]{Tho13}. These are measured using an adapted version of the Gas and Absorption Line Fitting code \citep[GANDALF v1.5;][]{Sar06} to derive emission line properties. GANDALF simultaneously fits stellar population and Gaussian emission line templates to the galaxy spectrum, in order to separate out the stellar continuum and absorption lines from the ionized gas emission. The effect of diffuse dust in the BOSS galaxies is taken into account assuming a \citet{Cal01} extinction law.

When deriving metallicities, we selected only host galaxies which were classified in SDSS as either `star-forming' or `composite', and which had measured fluxes in H$\alpha$, H$\beta$, [O~{\sc iii}]$\lambda$5007 and [N~{\sc ii}]$\lambda$6583; each with an amplitude-over-noise (AoN) of greater than 1.4. For lines which are detected below this threshold we set lower limits on the metallicity. As discussed previously, metallicities were not measured for AGN hosts.

After excluding AGNs, we obtain a sample of 581 SNe~Ia, of which 322 have AoN $>1.4$ for the lines needed to measure the metallicity. For the other 259 host galaxies we use the continuum flux in the region where emission lines are expected, to set an upper limit on their flux, and thus on the metallicity of the galaxy. We use the midpoint between this measured upper limit on the metallicity and the lower expected value for `normal' galaxies (12+log[O/H] = 7.10) as the estimate of the metallicity when testing for correlations and use the range between these two bounds as the error. Thus, these galaxies have extremely large error bars, they do not significantly affect the correlation fits, but are included for completeness. As a further test, these metallicity limits were excluded from the investigation of the correlations between SNe~Ia and host-galaxy parameters and consistent results were obtained (see Appendix~\ref{host_gal_no_lims}).

There are now emerging new (and improved) methods for measuring the metallicities of galaxies \citep{Kud14}. These stellar metallicities are based on low-resolution spectra of blue supergiant stars, using the such elements as iron, titanium, magnesium. However, as we are only concerned with ordering the host galaxies by their metallicities the absolute values are not so important.

We estimate the Star Formation Rate (SFR) for the galaxies using the H$\alpha$ line strength. We use H$\alpha$ as it is an intrinsically strong line and is located in the redder part of the spectrum, and so is less susceptible to dust extinction. The SFR estimates from the H$\alpha$ line are nearly instantaneous measures as the H$\alpha$ line is produced by ionizing photons which are generated by massive, young stars. We use the \citet{Ken98} relation to relate the H$\alpha$ luminosity to the SFR:

\begin{equation}
SFR=7.9\times 10^{-42}\times L(H\alpha) \mathrm{M}_{\odot} \mathrm{yr}^{-1}
\end{equation}
where $L(H\alpha)$ is measured in erg~s$^{-1}$. We measure the SFR for 523 non-AGN host galaxies, where the H$\alpha$ line is measured with AoN~$>$~1.4 (385 galaxies). The specific Star Formation Rate (sSFR) is a measure of the SFR in each host galaxy, scaled to the mass of the galaxy, i.e. per unit stellar mass. Additionally, for 58 galaxies where the flux was too low to actually measure the emission in H$\alpha$, the continuum was used to place an upper limit on the  H$\alpha$ emission, and hence set an upper limit on the SFR.

Table~\ref{sample_numbers} summarizes the sample size used in each analysis.

\begin{table}
\hspace{-0.8cm}
\begin{tabular}{cccc}
  \hline
Cut & Number kept  & Notes & Sample \\
  \hline
Full sample & 752 &  & \\
Fitted spectra & 602 & 150 removed &  \\
AGN `two-line' cut & 581 & 21 removed & Mass and age \\
H$\alpha$ AoN $>$ 1.4 &  581 & 523 values, 58 limits   & SFR and sSFR \\
All lines $>$ 1.4 & 581 &  332 values, 259 limits  & Metallicity \\
\hline
\end{tabular}
 \caption{Summary table of the sample sizes used in this analysis. The construction of each sample is described in Sect. \ref{host_properties}} 
\label{sample_numbers}
\end{table}

\section{SNe~Ia and host-galaxy properties distributions}
\label{sect:prop_dist}

\begin{figure}
\includegraphics[width=1\columnwidth,angle=0]{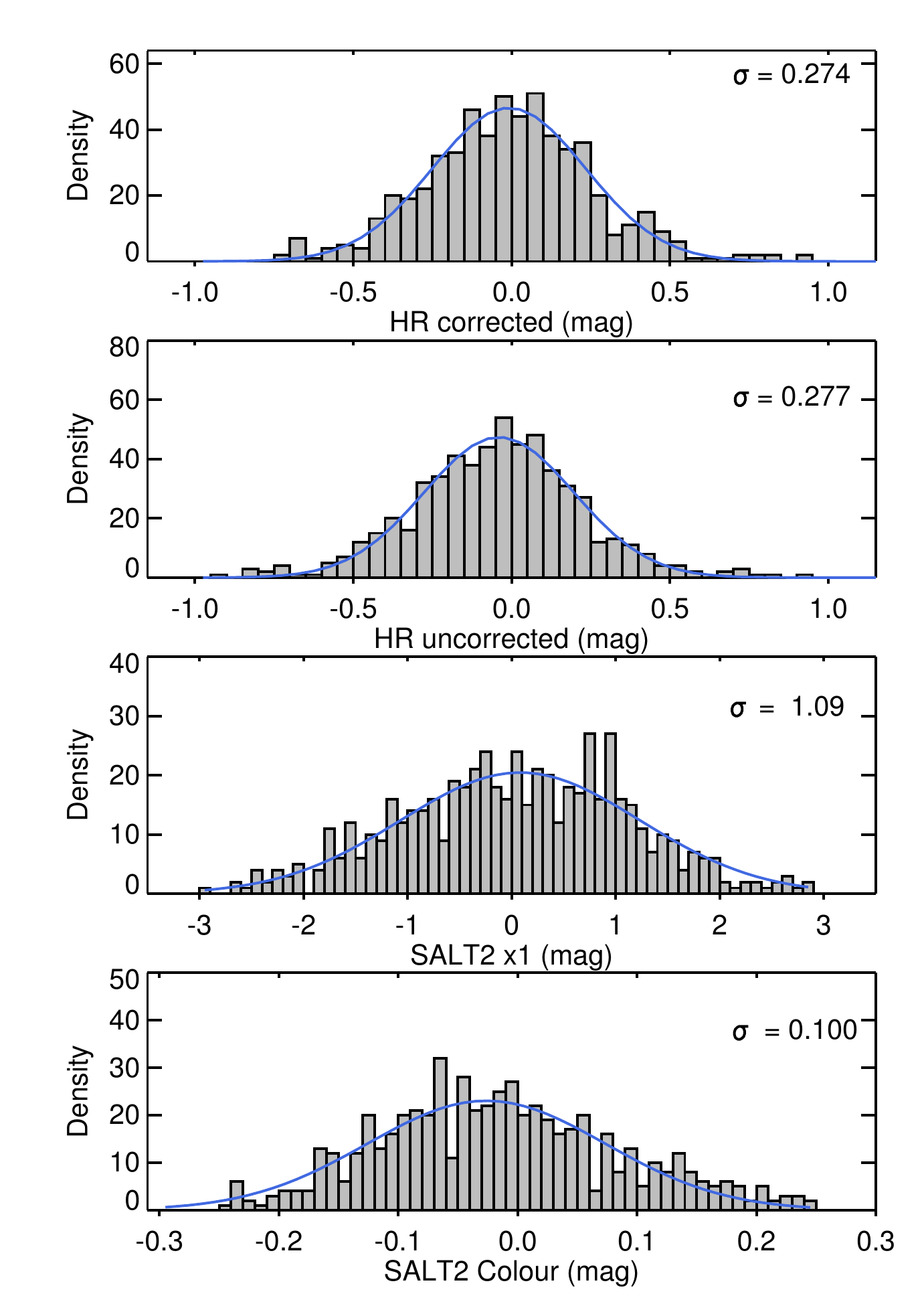}
\caption{\label{SN_parameters} One dimensional distributions of the SNe~Ia properties considered in this work. Gaussian fits to the histograms are over plotted in blue.}
\end{figure}

Fig.~\ref{SN_parameters} shows the distributions of the SNe~Ia parameters in our sample. The HR show a Gaussian distribution centred around zero, as might be expected from their definition as the residual for individual SNe from the best-fitting overall cosmology. The SALT2 $x_1$ parameter distribution has a skewness of only $-0.072$, i.e. slightly more bright SNe~Ia. This is to be expected in magnitude limited surveys, as at the limit of the survey brighter SNe~Ia will preferentially be observed. This Malmquist bias is corrected for within the cosmological analysis, although the effect is relatively small on the skewness of $x_1$. This is consistent with previous studies, such as \citet{Pan14}  and \citet{Rig13} who saw an even stronger bias to higher stretch SNe~Ia. The colour distribution has a larger skew in its distribution (skewness~=~0.28), and has a peak consistent with zero ($-0.03\pm0.1$) for most SNe~Ia but with a longer tail to redder colours. This is again consisted with previous studies \citep{Joh13,Rig13,Pan14}.

\begin{figure}

\includegraphics[width=1\columnwidth,angle=0]{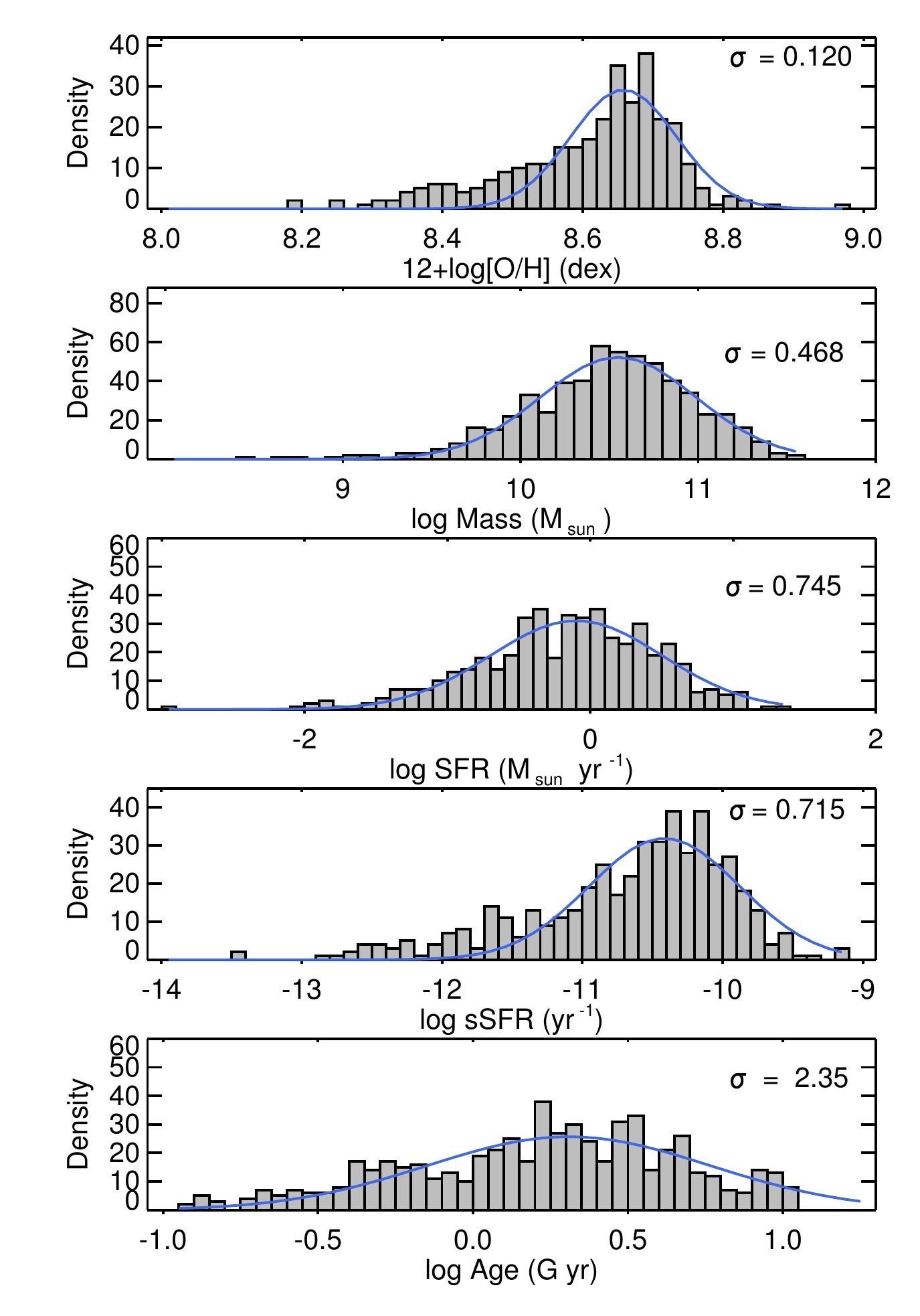}
\caption{\label{Host_parameters} One dimensional histograms of the SNe~Ia host-galaxy properties. Gaussian fits to the histograms (excluding limits) are over plotted in blue.}
\end{figure}

Fig.~\ref{Host_parameters} shows the distributions of SNe~Ia host-galaxy parameters in our sample.
The distribution of measured metallicities (i.e. excluding upper limits) has a skewness of $-0.90$ and an excess kurtosis of 0.94. The host-galaxy stellar mass distribution appears to be similar to previous SNe~Ia host galaxy studies \citep{Kel10,Rig13,Pan14}, with a skewness of $-0.62$ and an excess kurtosis  of 1.04. However, we seem to lack the lower mass host-galaxy population seen by some studies \citep{Lam10a,Chi13a}. The cause of this apparent difference is unclear, but may be due to the methods chosen for calculating the stellar mass, for example \citet{Lam10a} used the PEGASE2 code \citep{Fio97,Fio99} to calculate the stellar mass. Our host-galaxy mass distribution is consistent with the low redshift SDSS galaxy distribution \citep{Li09}. 

The SFR distribution is relatively Gaussian, with a peak at $-0.18\pm0.70$~M$_{\odot}$~yr$^{-1}$. The peak of the SFR distribution is shifted with respect to a comparable galaxy sample from photometric SFRs for the MPA-JHU SDSS catalogue \citep{Bri04} and the SNFactory SN host analysis \citep{Chi13b}. Our sample appears to lack the high SFR host galaxies seen by \citet{Bri04} and \citet{Chi13b}, however, a direct comparison is difficult as these other analyses use photometric estimates of the SFR rather than that measured from the H$\alpha$ emission. The distribution of sSFR  has a skewness of $-0.72$ and an excess kurtosis of 1.06. We also find a tail in the population of host galaxies which have lower SFRs, extending beyond the Gaussian envelope.

The logarithm of the age distribution is relatively Gaussian, with a skewness of 1.50 and an excess kurtosis of 1.88.  This is quite different to the age distribution of the BOSS galaxies \citep{Mar13}, which have a flat distribution with age. This shows that light in the majority of our SNe~Ia host galaxies is dominated by young stellar populations, although again there is a tail comprising of galaxies with ages up to 11 Gyr.

\section{Correlations between SNe~Ia and host-galaxy properties}
\label{sect:results}

To test for correlations between SNe~Ia parameters and host-galaxy properties, we employed a Bayesian linear regression technique (fitting a function of form $y=mx+c$), using the {\sc linmix\_err} package \citep{Kel07} for {\sc idl}. This method derives a likelihood function for the data being investigated using a Markov Chain Monte Carlo (MCMC) algorithm. The model accounts for measurement error in both parameters in the linear regression, and intrinsic scatter in the regression relationship. The technique outperforms other common estimators, and has been shown to be robust even when the measurement errors dominate the observed scatter, or when the distribution of independent variables are not Gaussian. 
As we are fitting data with large error bars, we set the METRO = 1 flag in {\sc linmix\_err}, so that the Metropolis-Hastings algorithm is used rather than the default Gibbs sampler, as this is helps when the measurement errors dominate the scatter in $x$ and $y$. As mentioned previously, when we have upper limits we use the midpoint between the lower expected value for galaxies (metallicity = 7.1 dex; SFR = $10^{-4} {\rm M}_\odot{\rm yr}^{-1}$) and the measured value, with the error bar stretching over the full range. We also repeat the analysis removing the limits, following the prescription for \citet{Kel07}. \citeauthor{Kel07} outline a method for including limits or censored data in  {\sc linmix\_err} in the dependent ($y$) variable. However, they suggest if the independent variable ($x$) is the limit then it is simpler to omit these limits, as inference on the regression parameters is unaffected
when a sample is selected based only on the independent variables. With this smaller sample with only measured parameters we find consistent correlations (see Appendix~\ref{host_gal_no_lims} for correlations excluding upper limits). 

The null hypothesis in our analysis is that there is no correlation between any of the SN parameters and the host galaxy parameters. We can reject this when a significant fraction of the MCMC samples are inconsistent with zero. The significant of the correlations is derived from the percentage of the posterior distribution which lies below zero (or vice versa for inverse correlations), while the uncertainty on the correlation is determined by from the 1$\sigma$ error on the Gaussian fit to the posterior distribution.

\begin{figure*}
\includegraphics[width=1.0\textwidth,natwidth=610,natheight=642,angle=0]{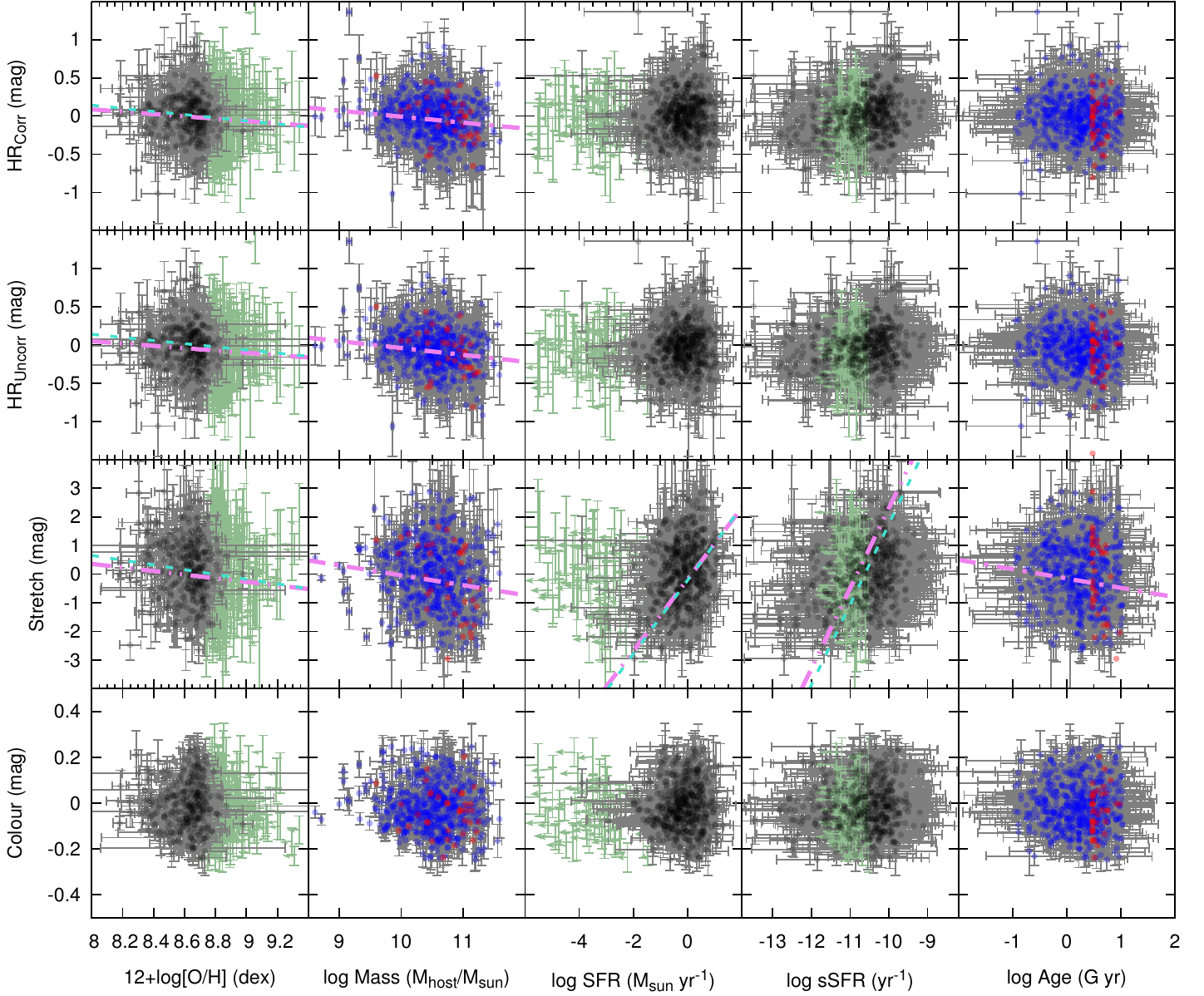}
\caption{\label{correlations} Correlations between SNe~Ia properties ($y$-axis) and host-galaxy properties ($x$-axis). Where a statistically significant slope was seen, the best-fitting linear relation is shown with a dot-dashed pink line. The slope for the metallicity correlations was not found to be significant, however since this is one of the correlations investigated by others we show it here for completeness. The dashed cyan line shows the best-fitting linear relation using only the measured parameters (excluding the limits) for the metallicity, SFR and sSFR. The blue and red points show the star forming and passive hosts, respectively, while the black points are used when the sample is not subdivided. The errors bars on the points are shown in grey, while the green arrows denote upper limits to values.}
\end{figure*}

The HR (both before and after correction for Malmquist bias, see \citet{Cam13} for details), SNe~Ia colour and $x_1$ value from the SALT2 fit (presented in \citealp{Cam13}) were compared to the host-galaxy metallicity, mass, age, SFR and sSFR.
The plots for all combinations of SNe~Ia and galaxy parameters are shown in Fig.~\ref{correlations}. The distributions are fitted with a line where we found a significant correlation. The best-fitting parameters and the significance are shown in Table~\ref{host_fits_2_line}.

By definition, if we take 20 samples from a Gaussian distribution, then $\sim$7 of them will lie $>1\sigma$ from the mean. As we are looking at 20 potential correlations between host galaxy and SN properties, we must be cautious of finding significant results for this reason. We hence adopt a 3$\sigma$ limit for our correlations; the chances of one of our twenty correlations being significant at this level by chance is $\sim$5 per cent. We note that in the following we find the correlation between host metallicity and HR to be {\it less} significant than this threshold, however, as a significant correlation has been found by other authors, it is still of interest to discuss here.

%When looking at 20 randomly-drawn Gaussian distributions, one would expect 6-7 to vary by more than 1$\sigma$ from the expected mean, and one of those to be more than 2$\sigma$. Here we find all but one vary by more than 1$\sigma$ and 10 which vary more than 2$\sigma$, suggesting that the corrections are not merely due to the nature of Gaussian distributions. All the posterior distribution except for (HR with sSFR) are $>1\sigma$ inconsistent with 0. 

%The posterior distributions of the correlations found from the {\sc linmix\_err} fit are shown in Fig.~\ref{significance}. 
\begin{table*}
\hspace{-1cm}
\centering
\begin{tabular}{cccccc}
 \hline
$x$ 		& $y$ 			& $m$ 												& $c$ (mag)						&sig 			& \% \\
\hline 
Metallicity	& HR corr       	& $-0.154\pm0.168$~mag/dex 							& $1.320\pm1.444$		& $1.4\sigma$ 	& 82\% \\ 
-       		& HR uncorr  	& $-0.152\pm0.165$~mag/dex 							& $1.267\pm1.420$		& $1.4\sigma$	& 82\% \\
-       		& $x_1$		& $-0.626\pm0.666$~mag/dex 							& $5.367\pm5.738$		& $1.4\sigma$ 	& 82\% \\
-       		& colour         	& $0.034\pm0.055$~mag/dex  							& $-0.308\pm0.478$		& $1.2\sigma$ 	& 73\% \\
\hline
Mass 	& HR            	& $-0.078\pm0.021$~\magpergalmass					& $0.772\pm0.211$		& $>5\sigma$ 	& 100\% \\
-        	& HR uncorr 	& $-0.090\pm0.021$~\magpergalmass					& $0.861\pm0.212$		& $>5\sigma$ 	& 100\% \\
-        	& $x_1$       	& $-0.347\pm0.068$~\magpergalmass					& $3.431\pm0.681$		& $>5\sigma$ 	& 100 \% \\
-        	& colour 	     	& $-0.011\pm0.006$~\magpergalmass 					& $0.081\pm0.065$		& $2\sigma$ 	& 95.83 \% \\
\hline
SFR 		& HR corr 		& $0.050\pm0.055$~mag/$\mathrm{log}({\rm M}_\odot{\rm yr}^{-1})$ 		& $-0.014\pm0.017$ 		& $1.4\sigma$  & 81.61\% \\
-  		& HR uncorr 	& $0.052\pm0.056$~mag/$\mathrm{log}({\rm M}_\odot{\rm yr}^{-1})$		& $-0.047\pm0.016$ 		& $1.4\sigma$  & 82.66\% \\
-     		& $x_1$ 		& $1.249\pm0.157$~mag/$\mathrm{log}({\rm M}_\odot{\rm yr}^{-1})$  		& $-0.184\pm0.051$ 		& $>5\sigma$   	& 100\% \\
-    		& colour 		& $0.036\pm0.019$~mag/$\mathrm{log}({\rm M}_\odot{\rm yr}^{-1})$  		& $-0.025\pm0.005$ 		& $2\sigma$   	& 95.99\% \\
\hline
sSFR 	& HR corr 		& $0.095\pm0.215$~mag/$\mathrm{log}({\rm yr}^{-1})$ 						& $1.015\pm2.308$  		& $0.9\sigma$	& 66.90\% \\
-  		& HR uncorr 	& $0.100\pm0.188$~mag/$\mathrm{log}({\rm yr}^{-1})$  						& $1.050\pm2.308$  		& $1.2\sigma$ 	& 70.03\% \\
-       		& $x_1$ 		& $2.802\pm0.762$~mag/$\mathrm{log}({\rm yr}^{-1})$   						& $30.324\pm8.149$ 	& $>5\sigma$   	& 100\% \\
-    		& colour 		& $0.547\pm0.226$~mag/$\mathrm{log}({\rm yr}^{-1})$  						& $5.866\pm5.602$ 		& $1.5\sigma$	& 86.55\% \\
\hline
Age 		& HR corr 		& $-0.059\pm0.032$~mag/$\mathrm{log}$(Gyr) 							& $0.004\pm0.015$ 		& $2.2\sigma$	& 97.25\% \\
 -      		& HR uncorr 	& $-0.068\pm0.034$~mag/$\mathrm{log}$(Gyr)  						& $-0.029\pm0.016$ 		& $2.3\sigma$ 	& 97.79\% \\
 -      		& $x_1$ 		& $ -0.327\pm0.106$~mag/$\mathrm{log}$(Gyr) 						& $-0.152\pm0.037$ 		& $3.7\sigma$	& 99.92\% \\   
 -      		& colour 		& $ -0.004\pm0.009$~mag/$\mathrm{log}$(Gyr)						& $-0.019\pm0.004$ 		& $1\sigma$   	& 69.79\% \\

  \hline    
\end{tabular}
 \caption{\label{host_fits_2_line} Summary of the fits and significance of the correlations between the host-galaxy properties and the SNe~Ia parameters with AGN removed by the `two-line' diagnostic. $m$ is the slope of the correlation, $c$ is the intercept with the $y$-axis. The columns `sig' and `\%' show the significance of the correlation, both in units of $\sigma$ and in the percentage of samples from the posterior distribution of slopes which lie above or below zero.}
 
\end{table*}

\subsection{Metallicity}

The left column of Fig.~\ref{correlations} shows the potential correlations between the SNe~Ia parameters and the metallicity of the host galaxy. The correlations between host-galaxy metallicity and HR all have a significance of between 1.2$\sigma$ and 1.4$\sigma$, so these do not pass the 3$\sigma$ threshold adopted previously.
We see a slight correlation (with a low significance of $1.4\sigma$) between the host-galaxy metallicity and the HR, both with and without the Malmquist bias correction. The slope of the correlation for the corrected HR is $-0.154\pm0.168$ mag/dex. The direction of the trend is such that metal rich galaxies have slightly brighter SNe~Ia after they have been standardized using SALT2 $x_1$ and colour (i.e. they have a negative HR).

\begin{figure}
\includegraphics[width=2\columnwidth,angle=0]{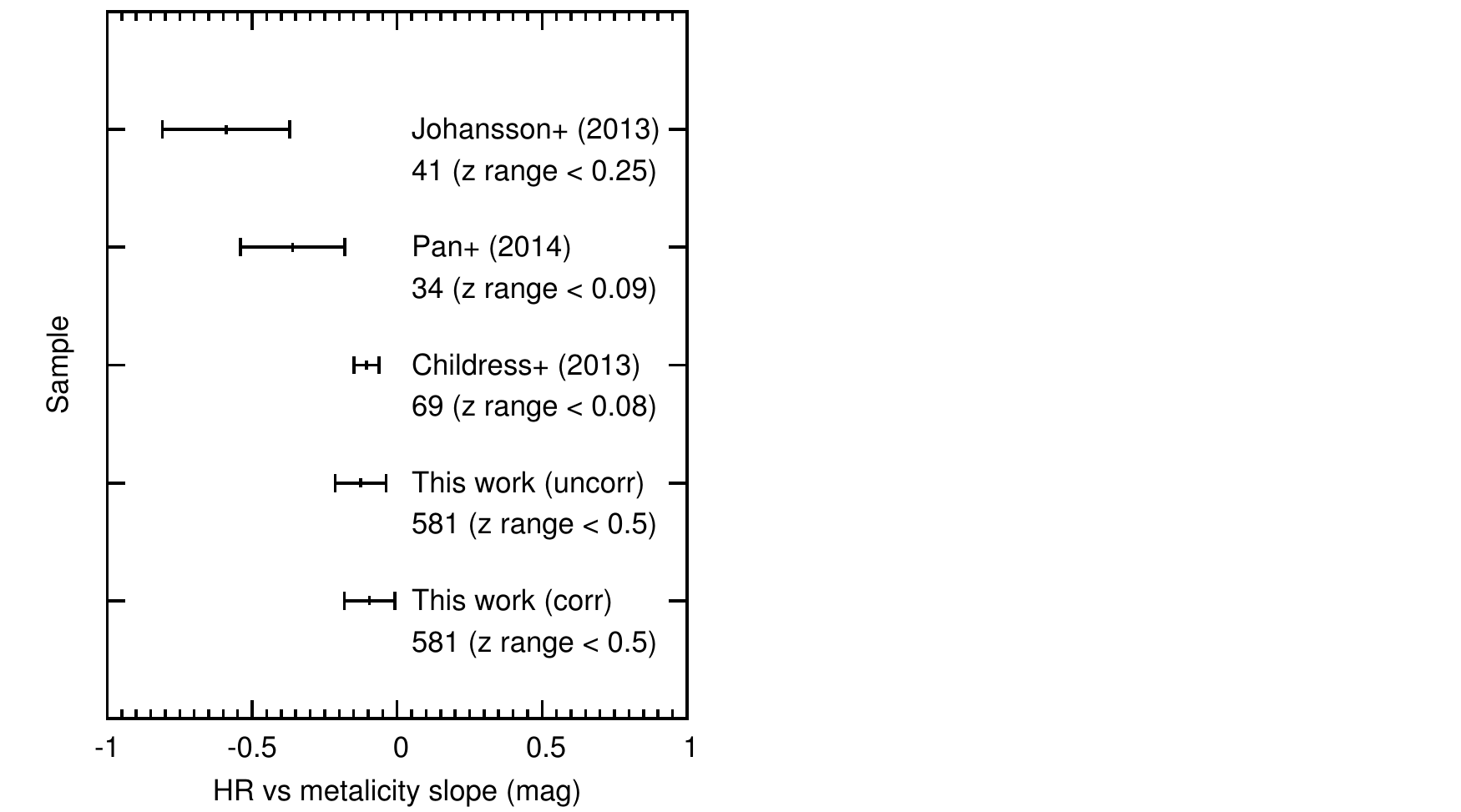}
\caption{\label{fig:comparison_HR_z} Slope of the correlation between HR and host-galaxy metallicity from different authors. Error bars correspond to the uncertainty in slope, the number beside each sample is the number of SNe~Ia from which the correlation was measured, and the redshift range of each study is also indicated.}
\end{figure}

This trend is in general agreement with previous studies \citep{DAn11,Kon11,Chi13a,Joh13,Pan14}. \citet{DAn11}, using 40 SDSS-II SNe~Ia in emission-line galaxies, found that light-curve corrected SNe~Ia are $\sim$0.1 magnitudes brighter in high-metallicity hosts than in low-metallicity hosts, at 4.9$\sigma$ significance. A comparison between the slope of the relation we find and that found by other authors is shown in Fig. \ref{fig:comparison_HR_z}. Our results appear to be consistent with all previous slopes apart from \citet{Joh13}, who see a steeper slope but do not find it to be statistically significant ($<2.5\sigma$). However, it is hard to directly compare our analysis to \citet{Joh13}, as they calculate metallicities from derived Lick indices. Our sample is 8 to 14 times large than these previous studies (when we include the limits from the continuum flux, or 4 to 5 times larger with only the measured values).

While we  do not see strong evidence for a correlation between the gas-phase metallicity of the host galaxy and either the SN stretch ($x_1$) or colour,  previous studies by \cite{Pan14} and \cite{Chi13a} saw a consistent correlation between these parameters. \citeauthor{Pan14} and \citeauthor{Chi13a} found that low-metallicity galaxies preferentially host broader (higher $x_1$) and redder (higher SALT2 colour values) SNe~Ia (before light-curve correction), with 98\% and 2.9$\sigma$ confidence, respectively. 
\citet{Chi13a} has considerably smaller errors bars, this may be due to the lower redshift range of the SNFactory sample used in their analysis ($0.03<z<0.08$) allowing for high-quality SN light curves to be obtained. In addition, the host-galaxy spectral follow up was carried out on 4--8m class telescopes, yielding high S/N spectra and hence more precise measurements of metallicities.  

It is perhaps surprising that a stronger correlation between the metallicity and SN HR is not observed. One might wonder if this is due to our low S/N data, however removing the limits and using only the sample below $z=0.3$ (which should have higher S/N spectra) does not significantly increase the significance of the correlation. We suggest that it might be that the integrated metallicity of the entire galaxy is not representative of the local environment where the SN progenitor formed, which may correlate stronger with the HR \citep{Rig15}.

\subsection{Mass}
\label{sect:mass}

\begin{figure}
\includegraphics[width=1.0\columnwidth,natwidth=610,natheight=642,angle=0]{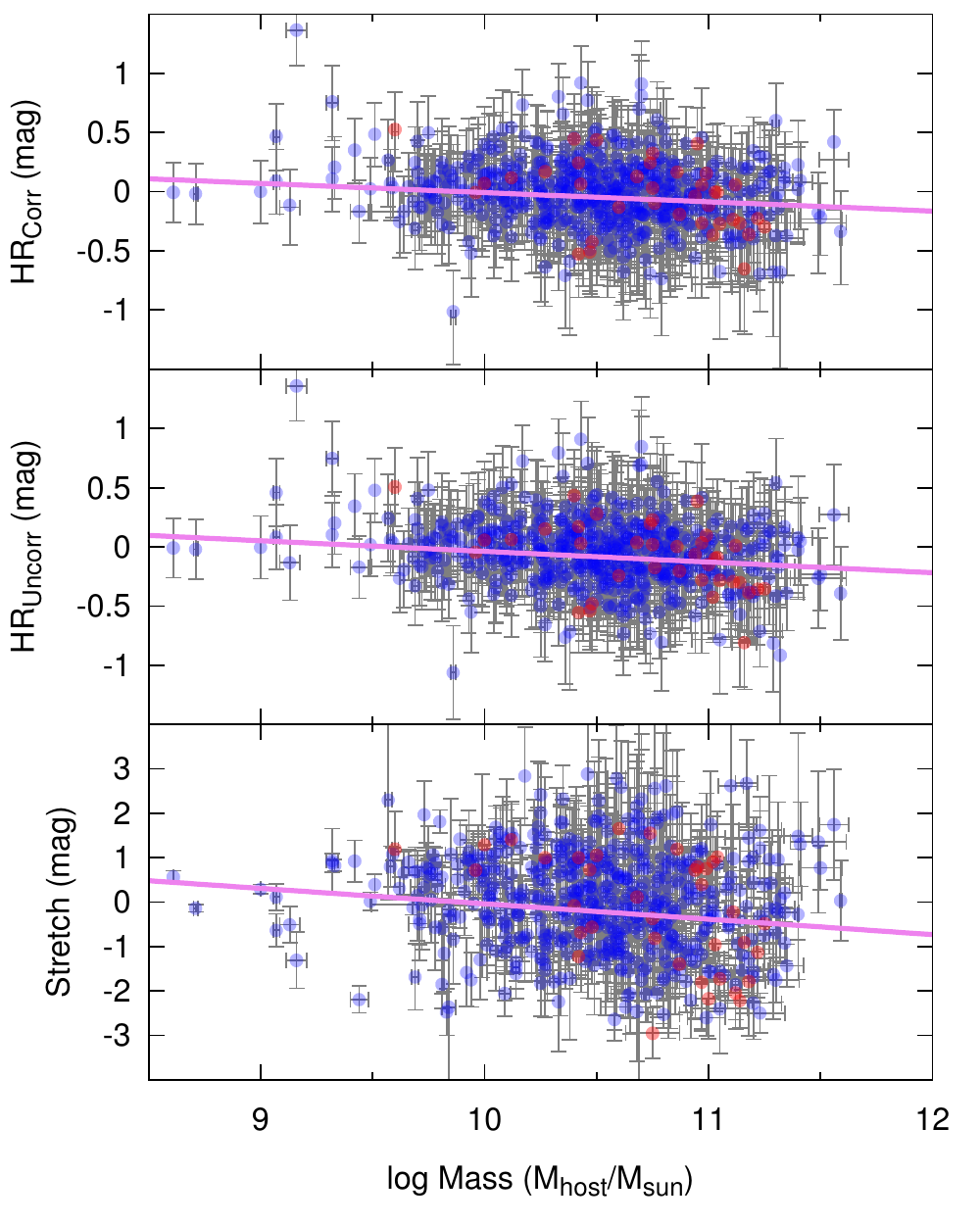}
\caption{\label{Mass_correlations} Correlations between SNe~Ia properties ($y$-axis) and host-galaxy mass ($x$-axis). The best linear fit to the data is shown with a pink line.  The blue and red points show star-forming and passive hosts respectively.}
\end{figure}

The second column of Fig.~\ref{correlations} shows the potential correlations between the SNe~Ia parameters and the stellar mass of the host galaxy. Fig.~\ref{Mass_correlations} shows magnified plots for all significant correlations between the host-galaxy mass and the SNe~Ia parameters. We see a highly significant ($>5\sigma$) correlation between the stellar mass of the host galaxy and both the Malmquist bias corrected and uncorrected HR; SNe~Ia which are brighter after light-curve correction preferentially explode in more massive hosts. We found the slope of this trend to be $-0.078\pm0.021$~\magpergalmass, which is consistent with most previous works. There also appears to be an offset between the passive and star-forming host galaxies, with more massive passive host galaxies with negative HRs, discussed further in Section~\ref{mass_met}. This is consistent with the idea that the mix of prompt and delayed channels varies between the passive and star-forming hosts, with the delayed channel dominating in passive host galaxies. It is surprising that the HR correlation with host-galaxy stellar mass  ($>5\sigma$) is much more significant than the correlation with metallicity ($1.4\sigma$). Although our sample of hosts with measured metallicity contains of only 332 galaxies (the other 259 having upper limits), a sample which is a factor of two smaller seems unlikely to account for a correlation which is a factor of six weaker. Indeed, if we test fitting the mass --  HR correlation with only the 332 galaxies with measured metallicity, we still find a $>5\sigma$ correlation, which actually has a steeper slope of $-0.117\pm0.031$~\magpergalmass\ (or a slope of $-0.134\pm0.030$~\magpergalmass when the HRs are uncorrected for Malmquist bias).

\begin{figure}
\includegraphics[width=1\columnwidth,angle=0]{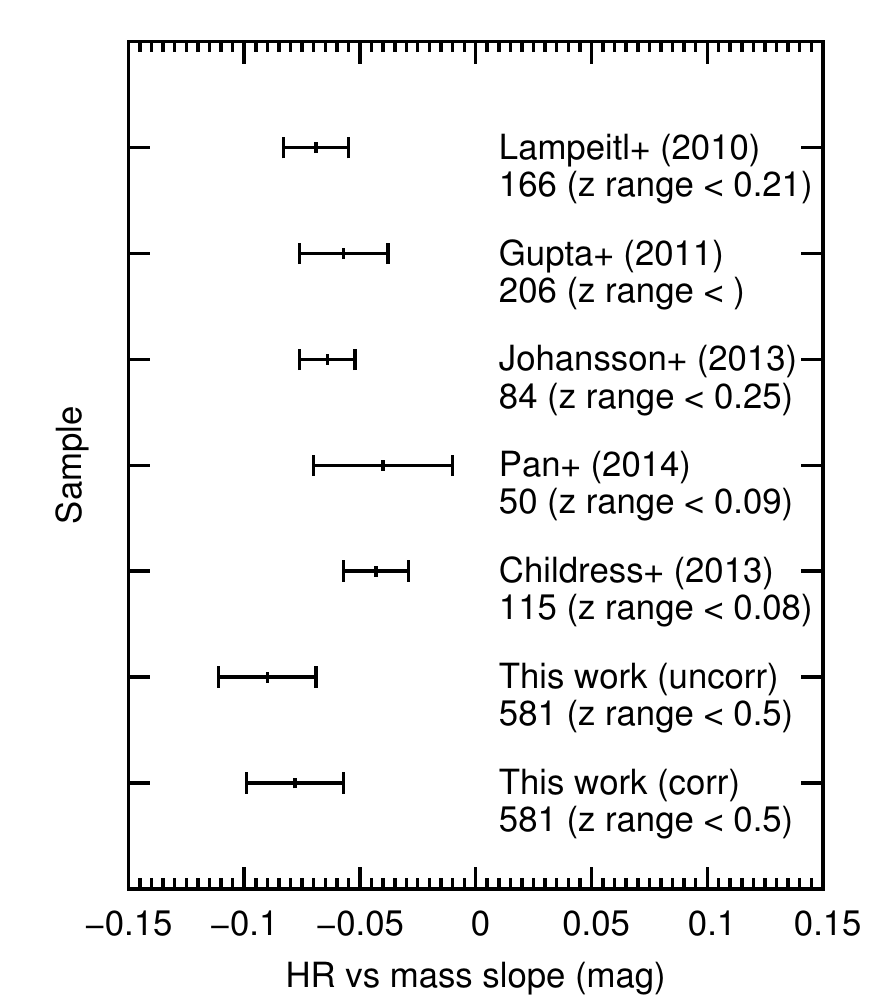}
\caption{\label{fig:comparison_HR_mass} Slope of the correlation between HR and host-galaxy stellar mass from different authors. Error bars correspond to the uncertainty in slope, the number beside each sample is the number of SNe~Ia from which the correlation was measured, the redshift range of each study is also given in parentheses.}
\end{figure}

Fig.~\ref{fig:comparison_HR_mass} shows a comparison of the host-galaxy mass versus HR slope from our work and that found in other analyses. Although we find a slope which is consistent with most previous studies, using our Malmquist bias corrected sample, our slope is slightly steeper than that found by \citet{Chi13a} using 115 SNe from SNFactory, who find a slope  of $-0.043\pm0.014$~\magpergalmass,
% \citet{Gup11} (206 SNe~Ia SDSS-II SN) find a slope of $-0.057\pm0.019~{\rm mag}/{\rm M}_{\rm sun}$, \citet{Lam10a} (116 SDSS-II SNe ~Ia) find a slope of  $-0.069\pm0.014~{\rm mag}/{\rm M}_{\rm sun}$, while \citet{Pan14} find a slope of $-0.04\pm0.03~{\rm mag}/{\rm M}_{\rm sun}$.

\begin{figure}
\hspace*{-0.9cm} 
\includegraphics[width=1.15\columnwidth]{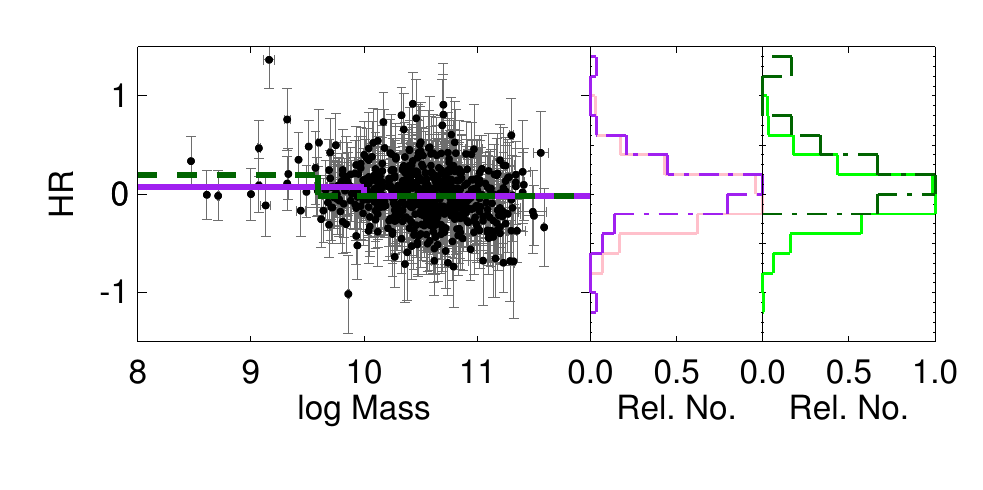}
\caption{\label{Mass_correlations_step} Left: correlations between SNe~Ia HR ($y$-axis) and host-galaxy mass ($x$-axis). The best-fitting step function to the data is shown with a purple solid line (split at \galmass~$=10$) and with the split as a free parameter in dark green dashed line (split at \galmass~$= 9.59$). Middle: normalised histograms of the SNe~Ia HR for high-mass (solid pink) and low-mass (dashed purple) split at \galmass~$=10$. Right: normalised histograms of the SNe~Ia HR for high-mass (solid green) and low-mass (dashed dark green) split at \galmass~$=9.59$}
\end{figure}

Other studies have split their sample of hosts into low- and high- mass galaxies, and fit each with a different value for the absolute magnitude of SNe~Ia, consistent with the direction of the trend we see in our data. We investigate dividing the sample into two subsamples, with a host-galaxy mass either above or below a threshold of \galmass$=10$, and fit a constant to each distribution of HRs using least squares. We then also investigate allowing the position of split between the two populations to vary as a free parameter. Fig.~\ref{Mass_correlations_step} shows these two fits to the data. We find that when the split is fixed at host-galaxy stellar mass of \galmass$=10$ the offset in HR between the two populations is $0.091\pm0.045$~mag, with a significance of 2.5$\sigma$. This offset between the two populations is comparable to that found in previous studies. \citet{Kel10} using a sample of 70 low redshift SDSS-II SNe~Ia found that physically larger, more massive hosts have SNe~Ia that are $\sim10$~per cent brighter after light-curve correction, which is an 0.11~mag offset in HR for SNe~Ia in low and high-mass hosts (which they define as log~M$>9.5~{\rm M}_{\rm sun}$) at 2.5$\sigma$ significance. \citet{Sul10}, using 195 SNe~Ia from SNLS showed that events of the same light-curve shape and colour are, on average, 0.08 mag ($\sim$4.0$\sigma$) brighter in massive host galaxies (which they define as \galmass$>10$) and galaxies with low sSFR. When the position of the split is allowed to vary, the best fit is found to be with a threshold of \galmass$=9.59$, and the offset in HR between the two populations is increased to 0.262~mag. While this is at a lower significance of only of 1.4$\sigma$, it is a larger HR offset than found in previous studies. With a freely varying mass threshold, the high mass subset has the same fit as when the split was forced to be at \galmass~$=10$. However, the low-mass population contains more positive HRs when the split between populations is allowed to vary, although it also consists of a smaller sample. There are only 18 SN below the \galmass$=9.59$ cut, so $\sim97$ per cent of the population are above the split, which in fact seems more consistent with no split. The histograms of HR for low- and high- mass galaxies using these two separations shown in Fig.~\ref{Mass_correlations_step} suggest that the high-mass galaxy sample contains more negative HRs in both cases. Only when the split is allowed to vary do we see a difference in the positive HR distribution, with lower-mass hosts tending to have SNe~Ia with a more positive HR, as found by previous studies \citep{Pan14}. However, we caution that as there are relatively few SNe~Ia (80; 14 per cent of the sample) with host masses below \galmass~$=10$, and the findings for SNe~Ia in low-mass hosts rely on small number statistics. To test the effects of this, we drew 80 galaxies at random to determine whether their mean HR differs from the remainder of the sample by more than the difference between the low- and high- mass subsamples discussed previously. Over 100 Monte Carlo iterations, 54\% of the instances we recover difference in mean HR between the samples as large as the 0.020 mag difference seen when allowing for a fixed mass cut. From this, we conclude that the difference between the high and low host galaxies is not likely to be significant.

In Fig.~\ref{correlations} and Fig.~\ref{Mass_correlations}, passive and star-forming galaxies are designated with red and blue points, respectively. As expected, the passive galaxies are more massive, and hence host the SNe~Ia with more negative HR. We also see a clear correlation between the stellar mass of the host galaxy and the SALT2 $x_1$ parameter at a significance of $>5\sigma$. SNe~Ia in more massive galaxies tend to have more negative $x_1$ (meaning they have narrower light curves prior to correction). The slope of the correlation between host-galaxy mass and  $x_1$  is  $-0.347\pm0.068$~\magpergalmass, which is in agreement  with previous studies \citep{How09,Nei09,Sul10,Chi13a,Pan14}. \citet{Joh13} see a stronger trend (4$\sigma$, in the same direction) between the host-galaxy mass and $x_1$, with a slope of $-0.75\pm0.19$~\magpergalmass. When determining this relation, \citet{Joh13} include AGN hosts in their sample, which may have some effect on the strength of the correlation.

Finally, we note that in comparing correlations with host-galaxy stellar mass between different studies, we must remain cognizant of the different techniques which were used to derive stellar mass. As discussed in Section~\ref{host_properties} we use the BOSS Portsmouth results, calculated using the \citet{Mar13} models, whereas some other previous studies, including \cite{Kel10} and \cite{Sul10} have used PEGASE2 for their stellar mass estimates. \citet{Chi13a} have considerably smaller errors bars, and this may be due to the wider wavelength range used to calculate the stellar masses (including UV through to IR photometry, rather than just optical photometry which we use here), as well as the lower redshift of their sample.

\subsection{Star Formation Rate} 

Columns 3 and 4 of Fig.~\ref{correlations} show the potential correlations between SNe~Ia parameters and the SFR and sSRF of the host galaxy. We find no significant correlation of HR with the SFR or sSFR of the host galaxy. While some other works \citep{Pan14} also saw no signs of a correlation between HR and SFR, others \citep{Sul10,DAn11,Chi13a} found a trend with sSFR with significance varying between 1.7$\sigma$ and 3.2$\sigma$, where host galaxies with lower sSFR tend to have over-luminous SNe~Ia after corrections (i.e. negative HR). \citet{DAn11} found a $3.1\sigma$ correlation, they included passive galaxies in their analysis (defined as having H$\alpha$ signal-to-noise$<$10 and failing one of their emission-line cuts). These slight differences we observe between the HR and the SFR or sSFR might be due to the population of low sSFR galaxies with positive HR which were absent in some previous samples \citep{DAn11,Chi13a}.

We see a highly significant correlation ($>5\sigma$) between $x_1$ and the SFR, with broader (higher $x_1$) SNe~Ia residing in galaxies with higher SFR. This correlation has a slope of $1.249\pm0.157$~mag/$\mathrm{log}({\rm M}_\odot{\rm yr}^{-1})$  for $x_1$ (or $2.802\pm0.762$~mag/$\mathrm{log}({\rm yr}^{-1})$ for $x_1$ with sSFR). This agrees with the recent work from \citet{Rig13}, where they measure the local star formation and observe that the previously noted correlation between stretch and host mass is driven entirely by the SNe~Ia coming from {\it locally} passive environments, in particular at the low-stretch end (at 3.8$\sigma$). Our data are not suited to make an analysis of the local environment (see Section ~\ref{location}), however using the SFR of the entire host galaxy we see the same trends, particularly at the low stretch end. Additionally, redder SNe~Ia appear to reside in galaxies with higher SFR. This correlation has a slope of $0.036\pm0.019$~mag/$\mathrm{log}({\rm M}_\odot{\rm yr}^{-1})$ (or $0.547\pm0.226$~mag/$\mathrm{log}({\rm yr}^{-1})$  with sSFR). This is again consistent with \citet{Rig13}, who found that SNe~Ia with local H$\alpha$ emission are redder by $0.036\pm0.017$ mag. The correlation with $x_1$ is also consistent with other previous studies e.g. \citet{Sul10}, who saw a $2.5\sigma$ difference between low and high sSFR and $x_1$. The correlation with colour is as one might perhaps expect, as star-forming galaxies contain more dust, and thus some of the SNe~Ia colour may be accounted for by host-galaxy reddening. This is in agreement with \citet{Pan14} who see the same correlation (at 3.1$\sigma$) with SNe~Ia colour. The $x_1$ and sSFR, and colour and sSFR are the only correlations to have highly skewed (1.331 and $-1.137$ respectively) and high kurtosis  (3.039 and 1.847, respectively) posterior distributions for the slope of the correlation.

\cite{Rig15} confirmed that that SNe~Ia in locally star-forming environments are dimmer than SNe Ia located in locally passive environments using the Constitution sample \citep{Hic09} and host-galaxy data from {\it GALEX}. They show that using samples with different distributions of locally passive and star-forming environments for the determination of $H_0$ results in an over estimate. \citeauthor{Rig15} find the resulting corrected value of $H_0$ to be $70.6\pm2.6$ \kms~$\mathrm{Mpc}^{-1}$, consistent with estimates of $H_0$ from the cosmic microwave background \citep{Pla15}.

In contrast, \citet{Jon15} investigated SNe~Ia in locally star-forming environments using GALEX imaging of the hosts of SNe~Ia from SDSS-II, SNLS, Pan-STARRS and Supernova Factory and find little evidence that SNe~Ia in locally passive environments are brighter, after light-curve correction, than SNe~Ia in locally star-forming environments. These authors suggest that the reduction in the significance of potential correlations is due to larger sample size and cleaner SNe~Ia selection criteria used for the \citet{Bet14} and \citet{Rie11} samples.

\subsection{Age}

The final column of Fig.~\ref{correlations} shows the potential correlations between the SNe~Ia parameters and the log age of the host galaxy. The only significant correlation we see is between the age of the host galaxy and the SALT2 $x_1$ parameter of the SNe~Ia ($3.7\sigma$). Narrower (more negative $x_1$) SNe~Ia are preferentially found in older stellar populations (with a slope of $-0.327\pm0.106$~mag/$\mathrm{log}$(Gyr)). This is the same general trend as seen before by \citet{Pan14} and \citet{Joh13}. However, \citet{Joh13}  found a much steeper slope of $-1.88\pm0.27$ at $>6\sigma$. They see very few old galaxies with broader SNe~Ia (positive $x_1$) values, where as we see far more of these. This is likely due to us pushing out to higher redshift and thus sampling a larger volume (see Fig.~\ref{z_less_0_3}, where we restrict our sample to $z<0.3$, and find a steeper correlation).

\citet{Rig13} suggested that the relation between SNe~Ia stretch and host-galaxy stellar mass is mainly driven by age, as measured by local SFR. \citet{Rig13} use locally passive environments to show that this drives the $x_1$-mass correlation, and that SNe~Ia with $x_1<-1$ arise exclusively in massive galaxies (log$(M/\mathrm{M}_{\odot})>$10). This is inconsistent with our findings, where we see a clear (albeit small) population of SNe~Ia in low-mass host galaxies with $x_1<=-1$.

\subsection{Comparison with Wolf et al.}
\label{sect:wolf}

The results presented here are in general agreement with the correlations presented in \citet{Wol15}, who used a similar sample of photometrically-classified SNe~Ia from SDSS-II, but re-determined host-galaxy parameters, stellar parameters and use stellar masses presented in \citet{Sak14} which were calculated using Flexible
Stellar Population Synthesis \citep[FSPS;][]{Con09,Con10}. Both this work and \citeauthor{Wol15} see a strong correlation with host-galaxy stellar mass ($>5\sigma$ in this work and 3.6$\sigma$ in \citeauthor{Wol15}). \citeauthor{Wol15} find a smaller offset in HR ($-0.044\pm0.011$~mag rather than $-0.066\pm0.045$~mag) when correcting for host-galaxy mass using a step function. This is due to the different host galaxy masses adopted for the location of the step function. Both analyses found weak evidence for a correlation between HR and host-galaxy metallicity, at a  significance of 1.4$\sigma$ in this analysis and 1.7$\sigma$ in \citeauthor{Wol15}. Furthermore, both works find no strong evidence for a trend of HR with sSFR ($0.9\sigma$ in this work and 0.42$\sigma$ in \citeauthor{Wol15}). The agreement between \citeauthor{Wol15} and this work are encouraging, as it suggests that despite different techniques for measuring host-galaxy stellar parameters and masses, the derived correlations between SN and host properties are robust.

\section{Subsample analyses}
\label{sect:subsample}

\subsection{SNe~Ia at low redshift}

\label{sec:z_less_0_3}

\begin{figure*}
\includegraphics[width=2\columnwidth,natwidth=610,natheight=642,angle=0]{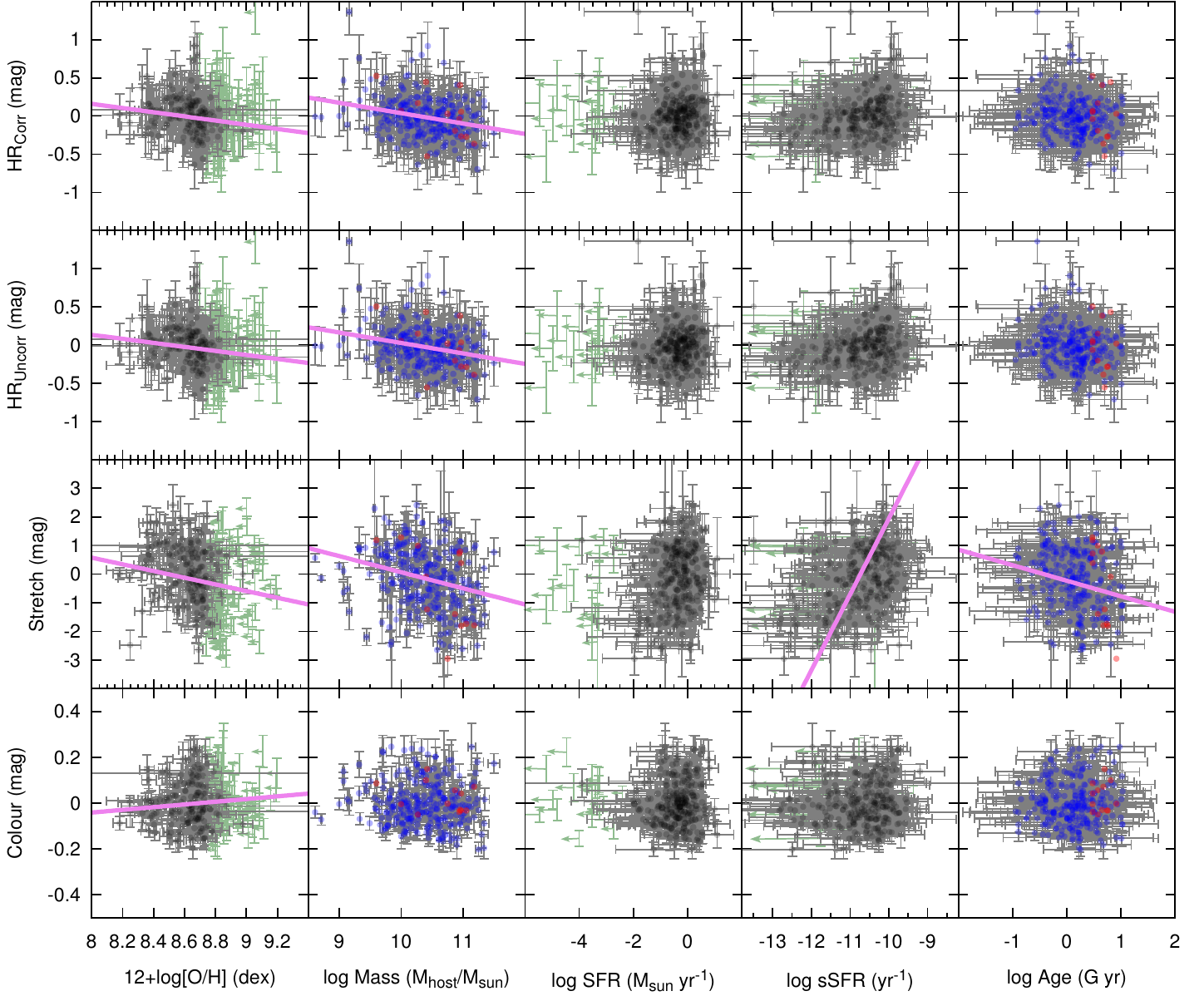}
\caption{\label{z_less_0_3} Correlations between SNe~Ia properties ($y$-axis) and host-galaxy properties ($x$-axis) for SNe~Ia with $z<0.3$. Where a statistically significant slope was seen in the best-fitting linear relation, it is shown with a pink line, as in Fig.~\ref{correlations}.  The slope for the metallicity correlations was not found to be significant, however since this is one of the correlations investigated by others we show it here for completeness. The blue and red points show the star forming and passive hosts, respectively, while the black points are used when there are only star-forming hosts in the plot. Green arrows demote upper limits to metallicity or SFR.}
\end{figure*}

\begin{table*}
\centering
\begin{tabular}{cccccc}
  \hline
$x$ 		& $y$ 			& $m$ 											& $c$ (mag)					& sig 			& \% \\
   \hline
Met		& HR corr 		& $-0.274\pm0.204$~mag/dex 						& $2.352\pm1.761$ 		& 1.7$\sigma$   & 90.81\% \\
  		& HR uncorr 	& $-0.264\pm0.206$~mag/dex 						& $2.247\pm1.774$ 		& 1.7$\sigma$   & 90.79\% \\
       		& $x_1$	 	& $-1.169\pm0.797$~mag/dex 						& $9.930\pm6.856$ 		& 1.8$\sigma$   & 91.19\% \\
   		& colour 		& $0.059\pm0.061$~mag/dex 						& $-0.513\pm0.528$ 		& 1.4$\sigma$   & 83.01\% \\
\hline
Mass	& HR corr 		& $-0.136\pm0.035$~\magpergalmass				& $1.398\pm0.363$		& $>5\sigma$   	& 100\% \\
  		& HR uncorr 	& $-0.137\pm0.034$~\magpergalmass 				& $1.398\pm0.354$		& $>5\sigma$   	& 100\% \\
       		& $x_1$ 		& $-0.563\pm0.110$~\magpergalmass  				& $5.698\pm1.145$		& $>5\sigma$   	& 100\% \\
    		& colour 		& $-0.013\pm0.010$~\magpergalmass   				& $0.135\pm0.107$		& $1.8\sigma$	& 91.20\% \\
\hline 
SFR 		& HR corr 		& $0.105\pm 0.103$~mag/$\mathrm{log}({\rm M}_\odot{\rm yr}^{-1})$ & $-0.018\pm0.024$ 		& $1.5\sigma$	& 84.74\% \\
  		& HR uncorr 	& $0.097\pm0.107$~mag/$\mathrm{log}({\rm M}_\odot{\rm yr}^{-1})$ 	& $-0.031\pm0.024$ 		& $1.4\sigma$	& 82.63\% \\
       		& $x_1$ 		& $1.798\pm0.377$~mag/$\mathrm{log}({\rm M}_\odot{\rm yr}^{-1})$ 	& $-0.306\pm0.114$  	& $2.9\sigma$ 	& 99.66\% \\
    		& colour 		& $0.031\pm0.045$~mag/$\mathrm{log}({\rm M}_\odot{\rm yr}^{-1})$ 	& $-0.008\pm0.008$  	& $1.2\sigma$	& 75.43\% \\
\hline
sSFR 	& HR corr 		& $0.097\pm0.226$~mag/$\mathrm{log}({\rm yr}^{-1})$  					& $1.047\pm2.440$  		& $0.9\sigma$	& 68.08\% \\
  		& HR uncorr 	& $0.097\pm0.233$~mag/$\mathrm{log}({\rm yr}^{-1})$ 					& $1.022\pm2.509$	 	& $0.85\sigma$	& 65.36\% \\
       		& $x_1$ 		& $2.645\pm0.638$~mag/$\mathrm{log}({\rm yr}^{-1})$ 					& $28.388\pm6.825$	  	& $>5\sigma$   	& 100\% \\
   		& colour 		& $0.505\pm0.179$~mag/$\mathrm{log}({\rm yr}^{-1})$ 				& $5.431\pm1.919$	  	& $2.8\sigma$	& 99.44\% \\
 \hline 
Age 		& HR corr 		& $-0.104\pm0.040$~mag/$\mathrm{log}$(Gyr)  					& $0.009\pm0.019$	 	& $2.8\sigma$  & 99.43\% \\
  		& HR uncorr 	& $-0.101\pm0.042$~mag/$\mathrm{log}$(Gyr)  					& $-0.008\pm0.019$	  	& $2.7\sigma$ 	& 99.12\% \\
       		& $x_1$ 		& $-0.543\pm0.131$~mag/$\mathrm{log}$(Gyr)  					& $-0.229\pm0.056$	   	& $>5\sigma$  	& 100\% \\
    		& colour 		& $0.005\pm0.001$~mag/$\mathrm{log}$(Gyr)  						& $-0.005\pm0.005$	  	& $0.85\sigma$	& 65.75\% \\    
  \hline    

\end{tabular}
 \caption{\label{z_less_0_3_table} 
 Summary of the fits and significance of the correlations between the host-galaxy properties and the SNe~Ia parameters with AGN removed by `two-line' diagnostic and with the sample restricted to SNe~Ia with redshift $z<0.3$. $m$ is the slope of the correlation, $c$ is the intercept with the $y$-axis. The columns sig and \% show the significance of the correlation, both in units of $\sigma$ and in the percentage of samples from the posterior distribution of slopes which lie above or below zero.}
 
\end{table*}

We also tested for correlations between SNe~Ia and host-galaxy properties after restricting our sample to $z<0.3$, to ensure that the observed correlations between host galaxy and the SNe~Ia parameters are not driven by the Malmquist bias in our sample, and to search for any evolution of parameters with redshift which may affect cosmological analyses. The low redshift ($z<0.3$) sample consists of 288 SNe~Ia, all of which have measured masses and ages, 271 with measured host-galaxy SFR and sSFR (the other 17 having limits from the non-detection of H$\alpha$), and 172 with measured metallicity (the other 116 having limits from the continuum flux of the spectral lines). We tested for correlations using the same procedure as in Section \ref{sect:results}, and the resulting plot of host galaxy and SNe~Ia properties is shown in Fig. \ref{z_less_0_3} and the fitted parameters in Table~\ref{z_less_0_3_table}.

From a comparison of Figs. \ref{correlations} and \ref{z_less_0_3}, it is clear that most of the correlations remain consistent between the full sample and the subset of low-redshift SNe~Ia. The mass correlations are consistent with the full sample, although we see that the slope of the correlations tend to be steeper at low redshift. 
Interestingly, the correlations between SNe~Ia $x_1$ and host SFR is significantly less significant at lower redshifts (going from a $>5\sigma$ result in the full sample to only 2.9$\sigma$). This could be the result of sampling a smaller range of host galaxies in the smaller volume at lower redshift. The redshift range of the sample was shown clearly to have a large effect in \citet{Sul10}, where they compared a low-redshift subsample to their full volume, and found that the low-redshift sample contained few low SFR galaxies (these galaxies also had low mass and metallicity).
The slope of the log age versus $x_1$ correlation seems to be somewhat sleeper in the low redshift sample. This is consistent with previous studies \citep{Joh13}, as there is a population of old galaxies with broad SNe~Ia (high $x_1$) at higher redshift, which are absent in the low redshift sample.

\subsection{Host spectra taken at the location of the SN~Ia} 
\label{location}

\begin{figure}
\includegraphics[width=1\columnwidth,angle=0]{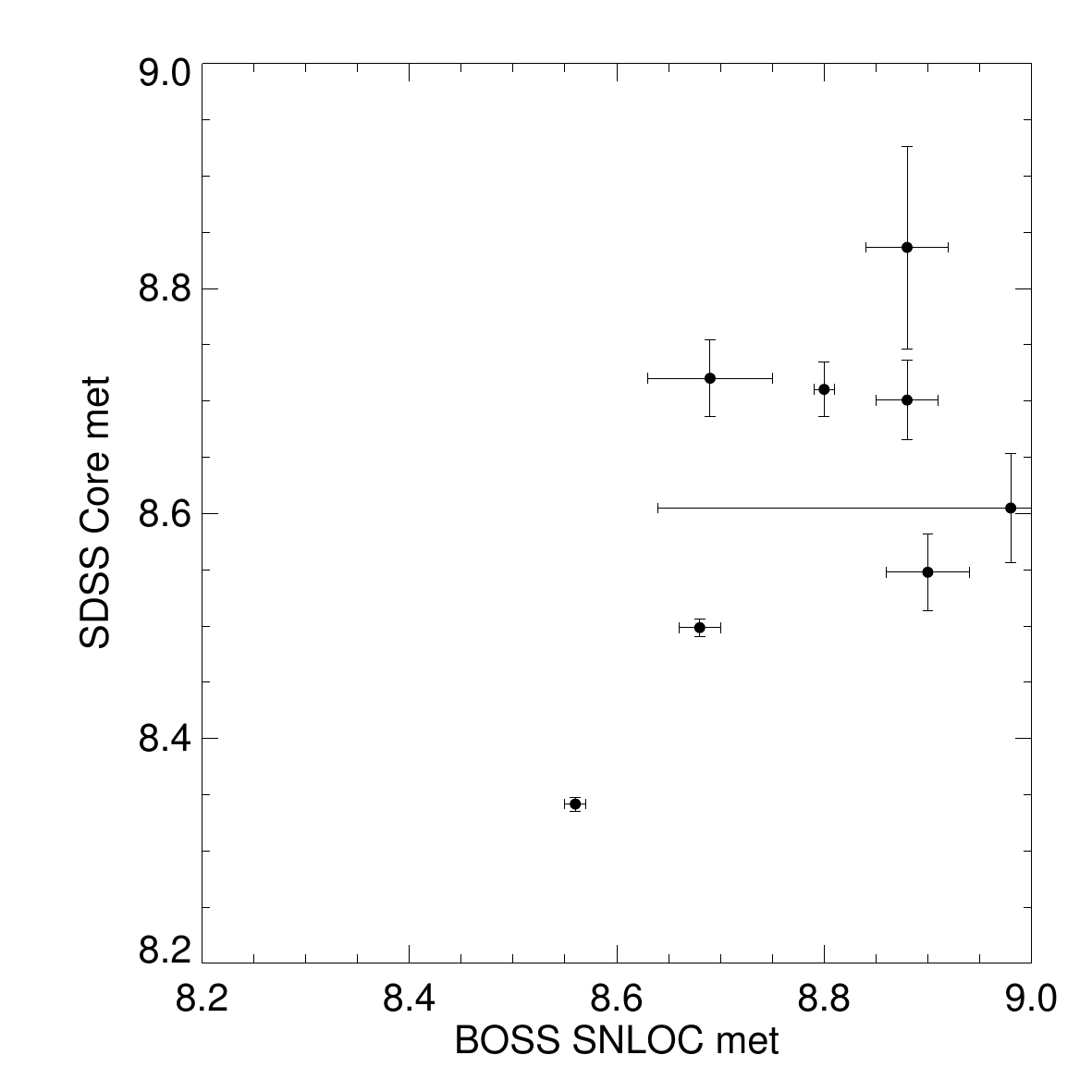}
\caption{\label{SDSSCore_SNLOC} This plot shows the metallicity for the host galaxies from the spectra taken at the core with SDSS compared to those taken at the position of the SN with BOSS. }
\end{figure}

For a small number of the SNe~Ia the BOSS spectra was taken at the position of the SNe~Ia rather than the core of the galaxy. This was mainly done for SNe~Ia where a host spectrum was already available in SDSS-II. From the sample of SNe~Ia with spectra at the SN position, we have 19 with measured host-galaxy properties, 13 of which are classed as star forming or composite, and 8 of which have AoN $>1.4$ for the lines used to measure metallicity.
 As these locations are in general well removed from the centre of the galaxy, we might expect to see different local properties than in the core of the host. For this sample we repeated the analysis for the full sample. It is hard to draw conclusions from such small sample, and the most significant correlation (between $x_1$ and host metallicity) was only at the $\sim2.2\sigma$ level.

In Fig.~\ref{SDSSCore_SNLOC} we compare the host-galaxy metallicities measured from spectra taken at the location of the SN to the metallicity derived from the SDSS spectra taken of the centre of the host. There is a systematic offset between the metallicity at the centre of the galaxy and at the SNe~Ia position, although the direction of the offset is opposite to that which we would expect given the typical metallicity gradient in galaxies. However, with a small sample and large error bars on metallicity measurements, along with the relatively large size of the SDSS and BOSS fibres, we caution that this offset is quite likely spurious.

\subsection{Separating the mass and metallicity correlations}
\label{mass_met}
\begin{figure*}
\centering
	\includegraphics[width=0.49\linewidth,height=0.59\linewidth,natwidth=610,natheight=642]{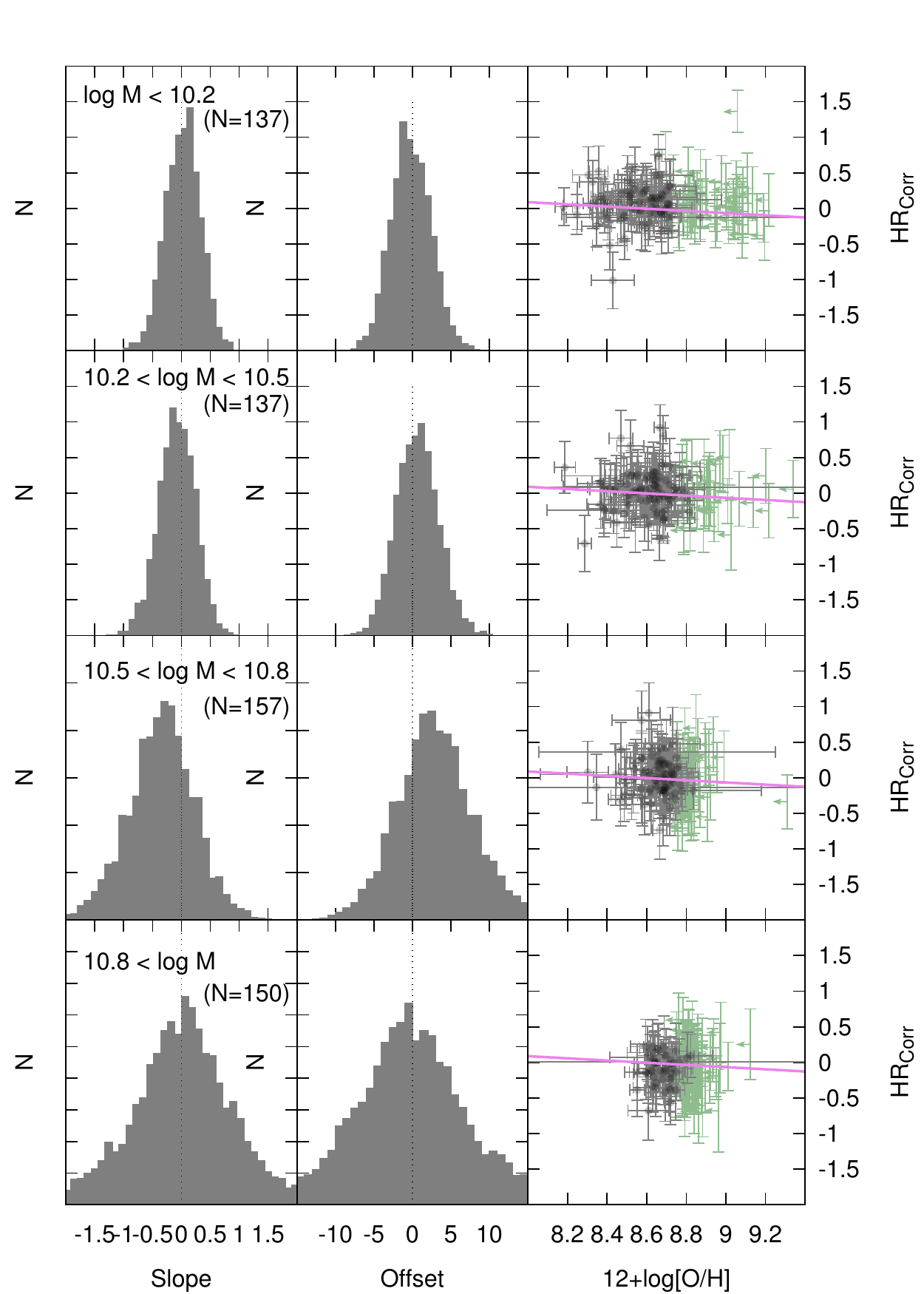}
	\includegraphics[width=0.49\linewidth,height=0.59\linewidth,natwidth=610,natheight=642]{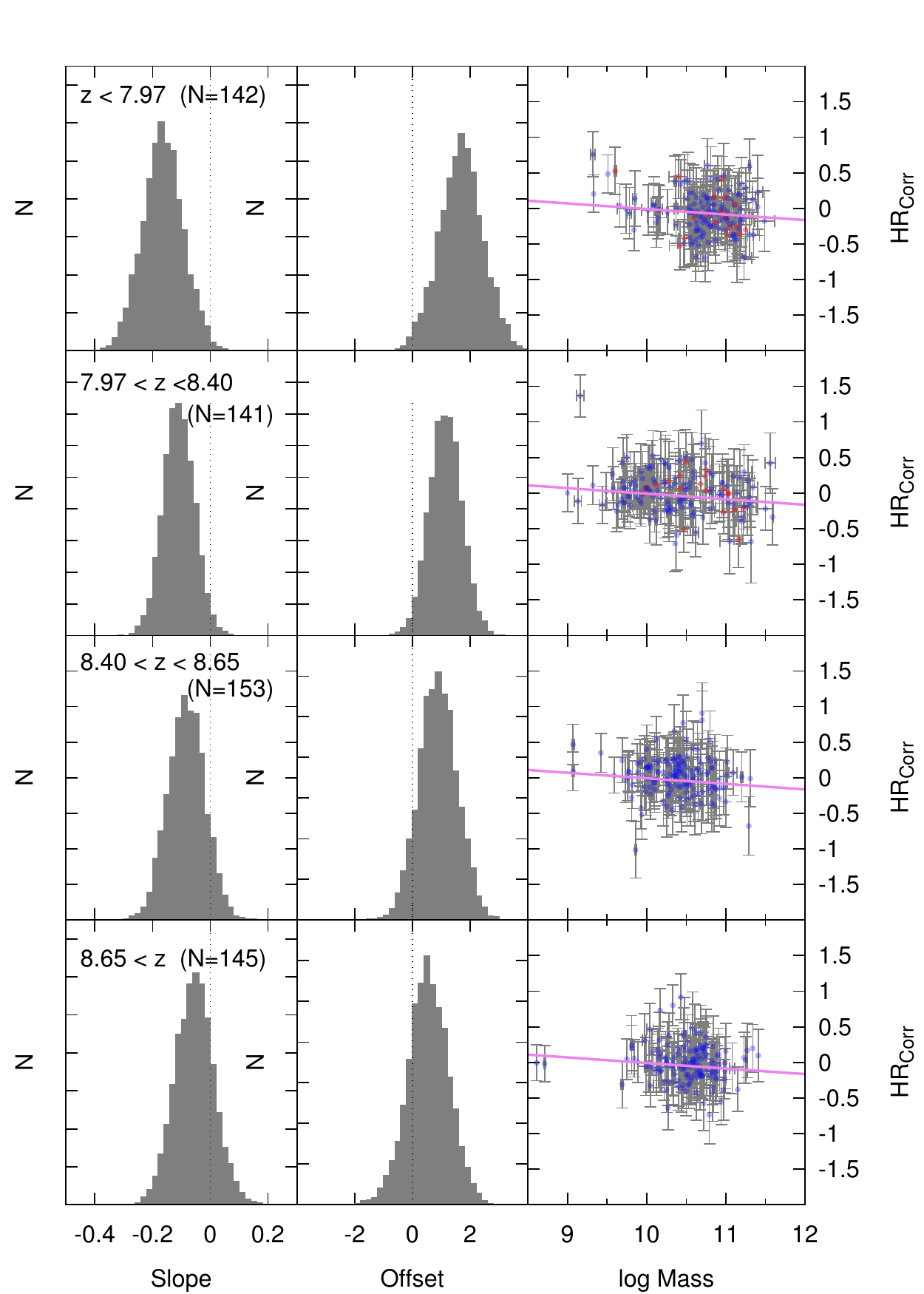}
\caption{\label{fig:bin} Left-hand panels: MCMC samples from the posterior distribution of slopes for the HR versus host-galaxy metallicity correlation. The sample has been divided into bins of increasing galaxy mass, and the distribution for each bin is shown separately. The first column shows the slope of the correlation ($m$), while the second column shows the offset ($c$). The third column shows metallicity against HR, along with the best-fitting correlation. The number of SNe~Ia in each mass bin is indicated in each panel.  Right-hand panels: the same plots, but for the HR versus host-galaxy mass correlation, as determined for samples in specific metallicity bins.}
\end{figure*}

The correlation between HR and host-galaxy metallicity and mass has been previously noted by many authors \citep{Sul10,Chi13a,Joh13,Pan14}. But what remains unclear is to what extent these correlations are related, as more massive galaxies are also more metal rich. To test this, we have divided our sample of SNe~Ia into bins corresponding to host-galaxy masses in a 0.3  dex range. We then looked for correlations between the SNe~Ia parameters and the host-galaxy metallicity {\it within each mass bin}, as was done for the larger sample. The results of this fitting within each bin are shown in Fig. \ref{fig:bin}. We also repeated this process, but binning the sample in metallicity, while searching for a correlation between HR and host-galaxy mass. As there is some scatter in the mass -- metallicity relation for galaxies, a sample of galaxies with the same mass will have a range of metallicities. By dividing the sample into bins of a given mass, or metallicity, we can control for the other variable and hence determine which is driving the correlation. If metallicity is the determining factor, then galaxies with similar mass, but different metallicities should show a correlation between metallicity and HR, while galaxies with a similar metallicity but different masses should not.

As can be seen in Fig. \ref{fig:bin}, the results of this test are somewhat ambiguous. For the two lowest galaxy mass bins, the peak of the posterior distribution corresponds to a slope of $\sim$0.0 (i.e. no correlation was found between metallicity and HR). While the $10.5<$~\galmass~$<10.8$ mass bin has a best-fitting correlation slope of $\sim-0.3$~\magpergalmass, there is a broad posterior distribution which is also consistent with zero slope. All the offsets are consistent with zero.
The highest mass bin has an even broader posterior distribution, which is peaked at zero. We note that as each of the mass bins only covers a relatively narrow range of metallicities, it is more difficult to measure the slopes of the host metallicity--HR relation than in the full sample.

However, when the sample was binned in metallicity, the posterior distribution of slopes for the HR with host mass correlation peaks at a negative slope for all of the bins, with the more metal poor hosts displaying a steeper slope, which seems to show the opposite to the trend we expect. The lowest metallicity hosts with $Z < 7.97$ have a best-fitting slope of $-0.16\pm0.07$~mag/dex, whereas the highest metallicity hosts with $Z > 8.65$ have a best-fitting slope of only $-0.06\pm0.06$~mag/dex. As the mass correlation is much stronger than the metallicity correlation in our full sample is it not surprising that the mass correlation still dominates in our binned distributions. It is unclear, however, why the slope of the host-mass relation is steeper in lower metallicity hosts. 

\begin{figure}
\includegraphics[width=1.0\columnwidth,natwidth=610,natheight=642,angle=0]{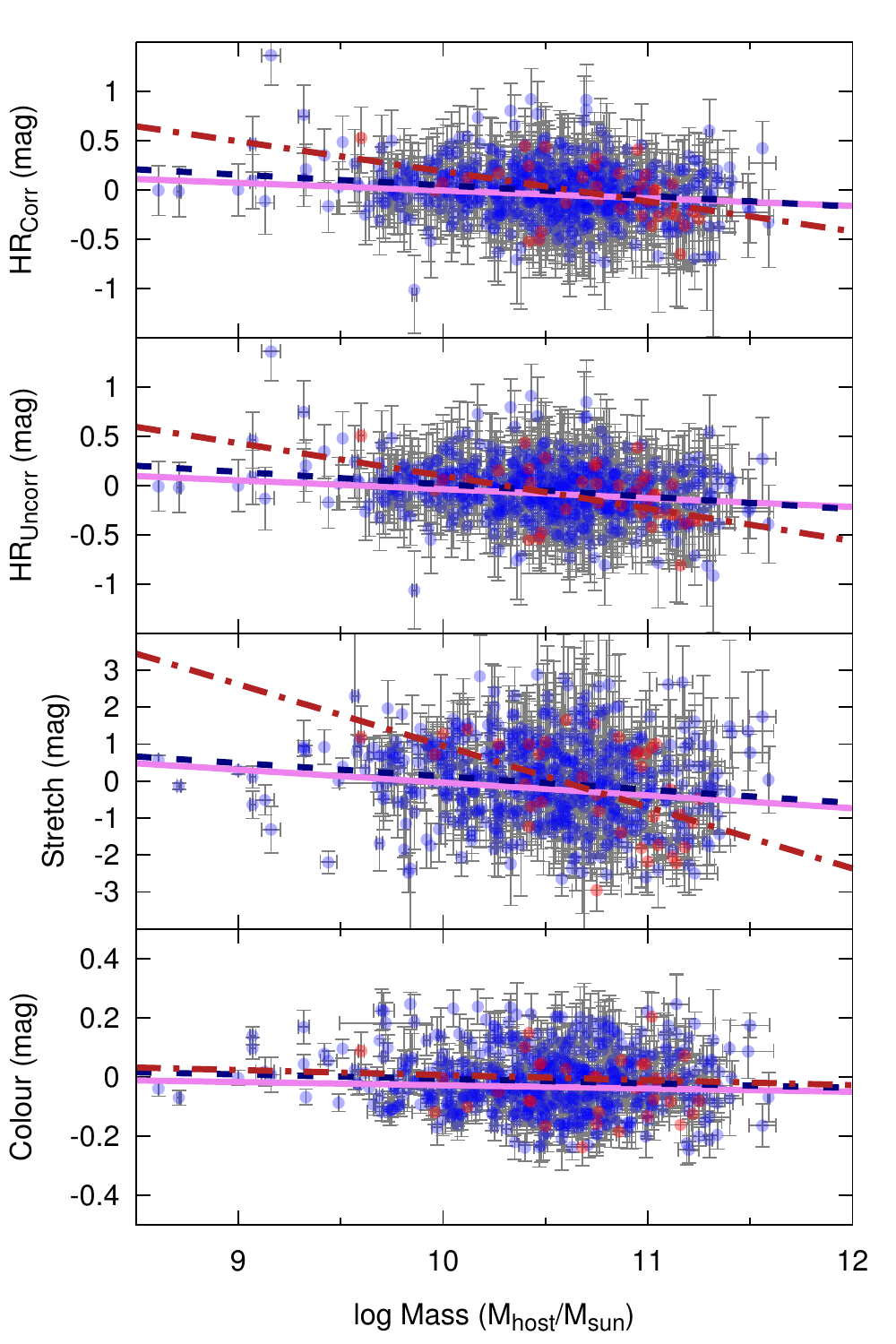}
\caption{\label{Mass_correlations_sf_pass} Correlations between SNe~Ia properties ($y$-axis) and host-galaxy mass ($x$-axis). The best linear fit to the combined data set is shown with a pink line.  The blue and red points show star forming and passive hosts, respectively. The best linear fit to the star-forming galaxies is shown with a blue dashed line and the best linear fit to the passive galaxies is shown with a red dot-dashed line.}
\end{figure}

The slope posterior distribution has a much more negative value in the lower metallicity bins. One possibility is that this may be due to the presents of passive galaxies in these lower metallicity bins driving the correlation, and there being no passive galaxies in the higher metallicity bins. To investigate this we look at the HR, stretch and colour for the passive and star forming galaxies separately, shown in Fig.~\ref{Mass_correlations_sf_pass}. This clearly shows that the passive galaxies have a much stronger dependence between the HR (and stretch) of the SNe~Ia and the host-galaxy mass than the star forming galaxies. The slopes are different at approximately the 1$\sigma$ level, as shown in the Table~\ref{SF_pass_table}. Using one universal correlation for all types of galaxies may end up under-correcting the SNe~Ia in passive galaxies. In Section~\ref{sect:cosmo} we investigate the effect that using separate correlations for SNe in star-forming or passive host galaxies has on the derived cosmological parameters.

\begin{table*}
\centering
\begin{tabular}{cccccc}
  \hline
$x$ 			& $y$ 			& $m$ (mag/dex)			& $c$ (mag)				& sig & \% \\

   \hline
SF mass   	& HR corr 		& $-0.108\pm0.030$		& $1.125\pm0.315$ 		& $>5\sigma$   & 100\% \\
  			& HR uncorr 	& $-0.126\pm0.029$		& $1.272\pm0.306$ 		& $>5\sigma$   & 100\% \\
       			& $x_1$ 		& $ -0.360\pm0.092$		& $3.728\pm0.964$ 		& $>5\sigma$   & 100\% \\ 
          		& colour   		& $-0.015\pm0.008$		& $0.143\pm0.088$ 		& 2.2$\sigma$   & 96.77\% \\
 \hline
  
Passive mass	& HR corr 		& $-0.305\pm0.160$		& $3.237\pm$1.717 		& 2.3$\sigma$   & 97.43\% \\
  			& HR uncorr 	& $-0.331\pm0.158$ 		& $3.408\pm$1.687 		& 2.4$\sigma$   & 98.00\% \\
       			& $x_1$ 		& $-1.661\pm0.506$ 		& $17.569\pm$5.399 	& 2.9$\sigma$   & 99.61\% \\ 
         		& colour   		& $ -0.017\pm0.042$		& $0.176\pm$ 0.453 		& 1$\sigma$   & 68.48\% \\
  \hline    

\end{tabular}
 \caption{\label{SF_pass_table} 
 Summary of the fits and significance of the correlations between the host-galaxy stellar mass and the SNe~Ia parameters. AGN have been removed by the `two-line' diagnostic, and the sample split divided into star-forming and passive hosts. $m$ is the slope of the correlation, $c$ is the intercept with the $y$-axis. The columns sig and \% show the significance of the correlation, both in units of $\sigma$ and in the percentage of samples from the posterior distribution of slopes which lie above or below zero.}
 
\end{table*}

\section{Cosmological analysis}
\label{sect:cosmo}

Previous studies such as ~\citet{Sul10} have shown that including a correction for the host-galaxy mass when carrying out cosmological analyses improves the quality of fit for the SNe~Ia data. The currently accepted paradigm in cosmology is the $\Lambda$~Cold Dark Matter ($\Lambda$CDM) model. This model uses a cosmological constant, $\Lambda$, to parametrize the  accelerating expansion of the Universe by dark energy, and is the simplest model we have to explain current observations. While the fiducial $\Lambda$CDM cosmological model has the dark energy equation of state parameter $w=-1$, if we allow this to be a free parameter we can test how close our Universe is to the $\Lambda$CDM cosmology.

We use the {\sc CosmoMC}  \citep{Lew02} code for our cosmological fits. We have used a flat $w$CDM cosmological model for fitting our data on its own, but allow $\Omega_\mathrm{k}$ ($\Omega_\mathrm{k}=1-\Omega_\Lambda + \Omega_\mathrm{m}$) to have values other than zero when fitting our data in combination with other cosmological information. The {\sc CosmoMC}  package uses an MCMC technique to efficiently probe multi-dimensional parameter space, allowing one to quickly investigate a large number of different regions in the parameter space. We allow simultaneous fitting of both the cosmological parameters and the SALT2 SNe~Ia parameters $\alpha$ and $\beta$, which allow for the standardization of SNe~Ia.  We also include in the distance modulus calculation the redshift-dependent Malmquist bias correction from \citet{Cam13} and the full SALT2 light-curve parameter covariance matrix. 

We ran the cosmological analysis on the sample of 581 SNe~Ia which have host-galaxy masses. We first carried out the fit on this sample as is, then again including a correction for the best-fitting linear correlation between host-galaxy mass and Malmquist bias-corrected HR (HR corr) as listed in Table~\ref{host_fits_2_line}. We also tested the effects of allowing for a linear correlation between HR and host mass within the cosmological fit, but allowing the slope and intercept parameters to vary freely, similarly to how the $\alpha$ and $\beta$ parameters of the distance modulus equation are treated. We see that the offset is not constrained in the cosmological analysis and is not correlated with any of the cosmological parameters, suggesting that having this extra degree of freedom is not required by the current data (see Appendix \ref{appendix_b} for more details). We hence rerun our cosmological analysis allowing only one free parameter to account for the host-galaxy mass. Additionally, we tested fitting this sample including a step-function in the relation between host-galaxy mass and Malmquist bias corrected HR (HR corr). We tested both the effects of a step functions fixed at \galmass$=10$ and with the best-fitting value of \galmass$=9.59$.

We investigated using the sample of SNe~Ia on their own, and with a prior on $H_0$ from the SH0ES analysis \citep{Rie11}. The SH0ES $H_0$ measurement is partially determined using nearby SNe~Ia measurements, and thus to be fully consistent we would have to consider the covariance between this value of $H_0$ and our SNe~Ia measurements. However, as we are assuming no prior information on $M$ in our treatment of intrinsic SNe~Ia parameters, these measurements can be considered independent. Using the SNe~Ia data with the SH0ES prior we fit for a flat, $w$CDM cosmological model using {\sc CosmoMC} and the prior Set I in Table~\ref{priors}. All other cosmological parameters are left at their default values at this stage.

Finally, we repeated the analysis combining our data with the power spectrum of Luminous Red Galaxies (LRGs) in the SDSS DR7 \citep{Rei10a}, and the full WMAP7 CMB power spectrum \citep{Lar11a}. We use WMAP data rather than the more recent Planck results \citep{Pla15} to facilitate direct comparison with previous studies. We fit this combination of data for a non-flat $w$CDM cosmology, using the priors listed as Set II in Table~\ref{priors}. With the addition of these external data sets, we can now relax our priors on the re-ionization optical depth ($\tau$=[0.00, 0.50]), the primordial super-horizon power in the curvature perturbation on 0.05 Mpc$^{-1}$ scales ($\mathrm{log}~A$=[0,30]), and the scalar spectral index ($n_s$=[0,1.5]), which had previously all been set to zero.

\begin{table}
\centering
\begin{tabular}{|c|c|c|}
  \hline
  Parameter  & Set I & Set II  \\ 
  \hline
  $w$ &  [-3,3] & [-3,3] \\
 $\Omega_{\rm k}$ &  0 & [-1.5,1.5] \\
  $\Omega_{\rm dm}$  & [0.0, 1.2] & [0.0, 1.2] \\ 
  $\Omega_{\rm b}$  & 0.0458  & [0.015, 0.200] \\
  $H_0$  & [50,100] & [50,100] \\

\hline
\end{tabular}
 \caption{Priors imposed on the fitted cosmological parameters in the two different combinations (sets).}
\label{priors}
\end{table}

\begin{figure*}
\centering
\includegraphics[width=1\linewidth,height=1\linewidth]{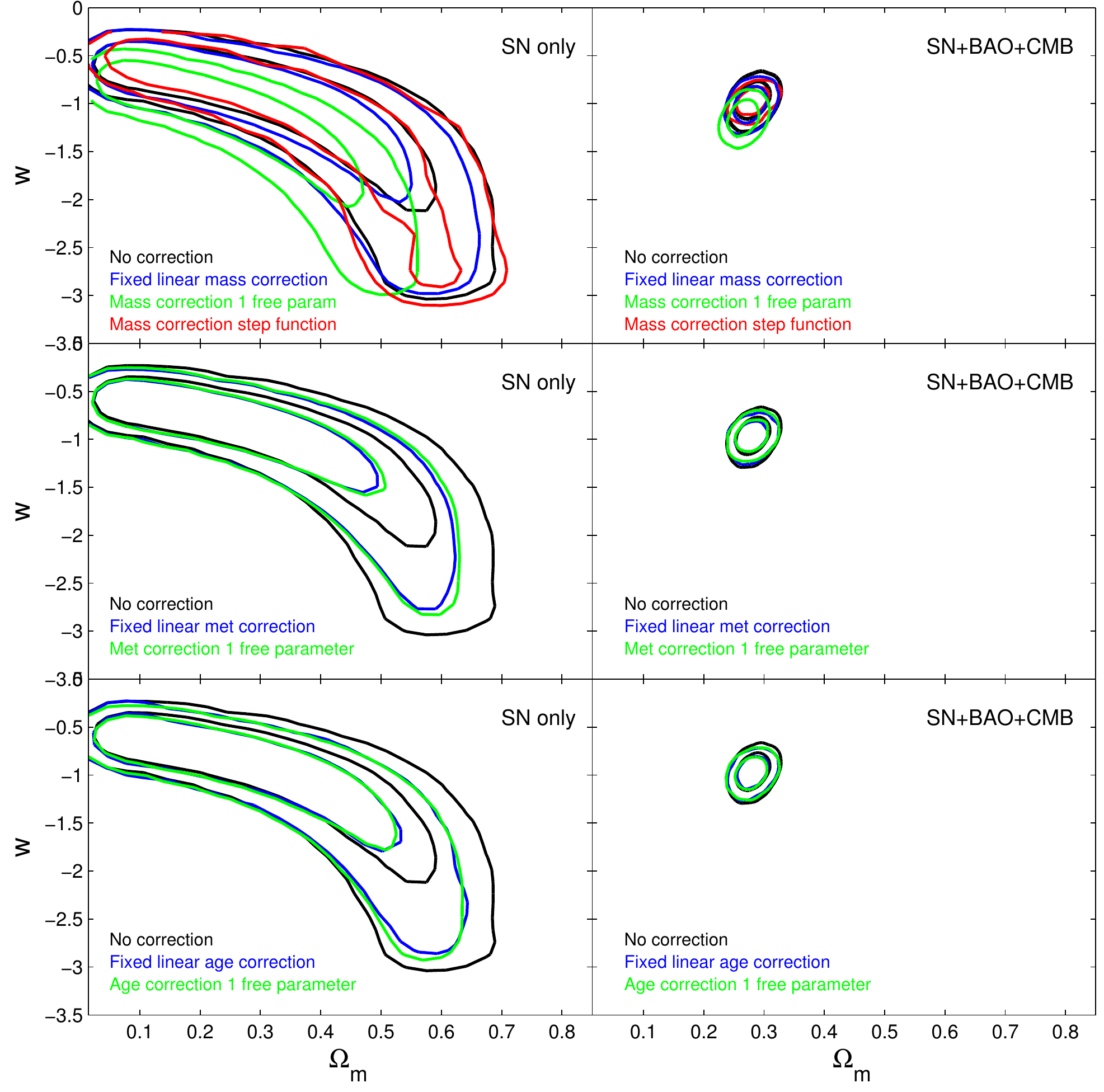}
\caption{\label{cosmology_mass}
$w$ versus $\Omega_\mathrm{m}$  for the sample of 581 SNe~Ia with measured host-galaxy properties. Left-hand panels: $w$ versus $\Omega_\mathrm{m}$  using only SNe~Ia data together with a prior on $H_0$. Right-hand panels: SNe~Ia~+~$H_0$~+~BAO~+~CMB. The black contours are uncorrected in all panels. Top panels: the blue contours are corrected for the host-galaxy stellar mass using the best-fitting linear function, with $m=-0.078$~\magpergalmass~and $c=0.772$~mag. The red contours are corrected for the host-galaxy metallicity using a step function split at a stellar mass of \galmass$=10$, with $0.091\pm0.045$~mag as the linear offset between the two bins. Middle panels: the blue contours are corrected for the host-galaxy stellar metallicity using the best-fitting linear function, with $m=-0.154$~mag/dex and $c=1.320$~mag. Bottom panels: the blue contours are corrected for the host-galaxy log age using the best-fitting linear function, with $m=-0.059$~mag/G~yr and $c=0.004$~mag. The red contours have $m$ and $c$ as free parameters in the {\sc CosmoMC} fit, the green contours have only $m$ as a free parameter for the mass, metallicity and age correlation in the top, middle and bottom panels, respectively. The contours enclose 1$\sigma$ and 2$\sigma$ limits on $w$ and $\Omega_\mathrm{m}$.}

\end{figure*}

The top-left panel of Fig. ~\ref{cosmology_mass} shows the contours for $\Omega_\mathrm{m}$ plotted against the dark energy equation of state parameter, $w$, for the sample of 581 SNe~Ia with measured host-galaxy stellar mass, using only SNe~Ia data plus the prior on $H_0$. The cosmological contours for all the potential correlations between parameters in the {\sc CosmoMC} fit can be found in the Appendix. The best-fitting cosmological parameters are listed in Table~\ref{cosmo_table}. When only using SNe~Ia data and the $H_0$ prior, applying a linear correction for the host-galaxy mass (either with the best-fitting values quoted in Table~\ref{host_fits_2_line} or with the coefficient left as a free parameter in the fit), the size of the cosmological contours are reduced, as shown in the top-left hand panel of Fig.~\ref{cosmology_mass}. When the parameters of the correction for the host-galaxy stellar mass are fixed in the cosmological analysis it biases the $\alpha$ and $\beta$ parameters, more details can be found in the Appendix. The contours also shift to more negative $w$ and lower values of $\Omega_\mathrm{m}$, especially when the parameters are left free. This suggests that without a host-galaxy correction for mass the cosmological contours are biased. The linear correlation of HR with host-galaxy mass is found to have a slope of $-0.123\pm0.021$ within the {\sc CosmoMC} fit with one free parameter. This slope is slightly steeper than that found when fitting the HR and mass after a cosmological solution had already been found (i.e. Section \ref{sect:mass}; $-0.078\pm0.021$), but is consistent with the slope from the low redshift sample in Section \ref{sec:z_less_0_3} ($-0.136\pm0.035$). When only using SNe~Ia data and the $H_0$ prior, applying a step-function correction for the host-galaxy mass, there is very little difference in the size or position of the cosmological contours. This is somewhat surprising as previous studies suggested that the step-function represented the data well. \citet{Bet14} using the SDSS+SNLS data found a 5$\sigma$ step function correction, with an offset of $-0.07\pm0.02$~mag between the high and low-mass host galaxies (with a split at \galmass$=10$). The much lower significance of the step function we find is likely due to the small numbers of SNe~Ia in hosts with masses below \galmass$=10$ in our data, compared to the sample of \citet{Bet14} which contained many more SNe~Ia in low-mass hosts.

The cosmological contours for $w$ versus $\Omega_\mathrm{m}$, after combining the SNe~Ia data with the power spectrum of LRGs in SDSS DR7 \citep{Rei10a} and the full WMAP7 CMB power spectrum \citep{Lar11a} are shown in the top right-hand side of Fig.~\ref{cosmology_mass}, with the best-fitting values quoted in Table~\ref{cosmo_table}. Again, the cosmological contours shift to slightly lower values of $w$ and smaller values for $\Omega_\mathrm{m}$ when the host-galaxy mass correlation is included. The change in the best-fitting value of $w$ is 0.18, which is a $\sim1\sigma$ difference. In fact, with one or two parameters free in the {\sc CosmoMC} the best-fitting cosmology excludes $w=-1$ at the 1$\sigma$ level. However, since this analysis is purely statistical, with no systematic errors taken into account this is not a significant result.

We repeated the cosmological analysis, but correcting the SNe~Ia distance moduli by the metallicity or the age of the host rather than the mass. In this case, we used the same size SNe~Ia sample (581) but with 259 (45 per cent) of them having limits rather than measured host-galaxy metallicities (all 581 host galaxies have measured ages). Again, we begin by including two free parameters for the correlation with host-galaxy metallicity or age, the slope ($m$) and offset ($c$). Similar to the host-galaxy stellar mass correlation, we see that the offset is not constrained at all, indicating that the data do not require a parameter for the offset as well as slope in the host-galaxy metallicity or age correlation. We thus performed the cosmological analysis with only one free parameter for the slope of the metallicity or age correlation.

The cosmological contours for all the potential correlations between parameters in the {\sc CosmoMC} fit, for the sample of 581 SNe~Ia with measured host-galaxy metallicity, using only SNe~Ia data plus the prior on $H_0$ and combing with other cosmological probes are shown in the Appendix. The $w$ versus $\Omega_\mathrm{m}$ cosmological contours from the metallicity correlation analysis are shown in middle row and the age correlation analysis are shown in the bottom row of Fig.~\ref{cosmology_mass} . The best-fitting cosmological parameters are given in Table~\ref{cosmo_table}. Again the left-hand panels show the fit for the sample using only SNe~Ia data with the $H_0$ SH0ES prior \citep{Rie11} and a flat $w$CDM cosmological model. The right-hand panels show the effect of also including the LRGs \citep{Rei10a}, and the full WMAP7 CMB power spectrum \citep{Lar11a}. This clearly shows that including a correction for the host-galaxy metallicity or age reduces the size of the error contours when only SNe~Ia data and a prior on $H_0$ are used; but when other cosmological probes are included in the analysis the effect of the host-galaxy metallicity or age correlation is negligible.

\begin{figure*}
\centering	
	\includegraphics[width=0.45\linewidth,height=0.45\linewidth]{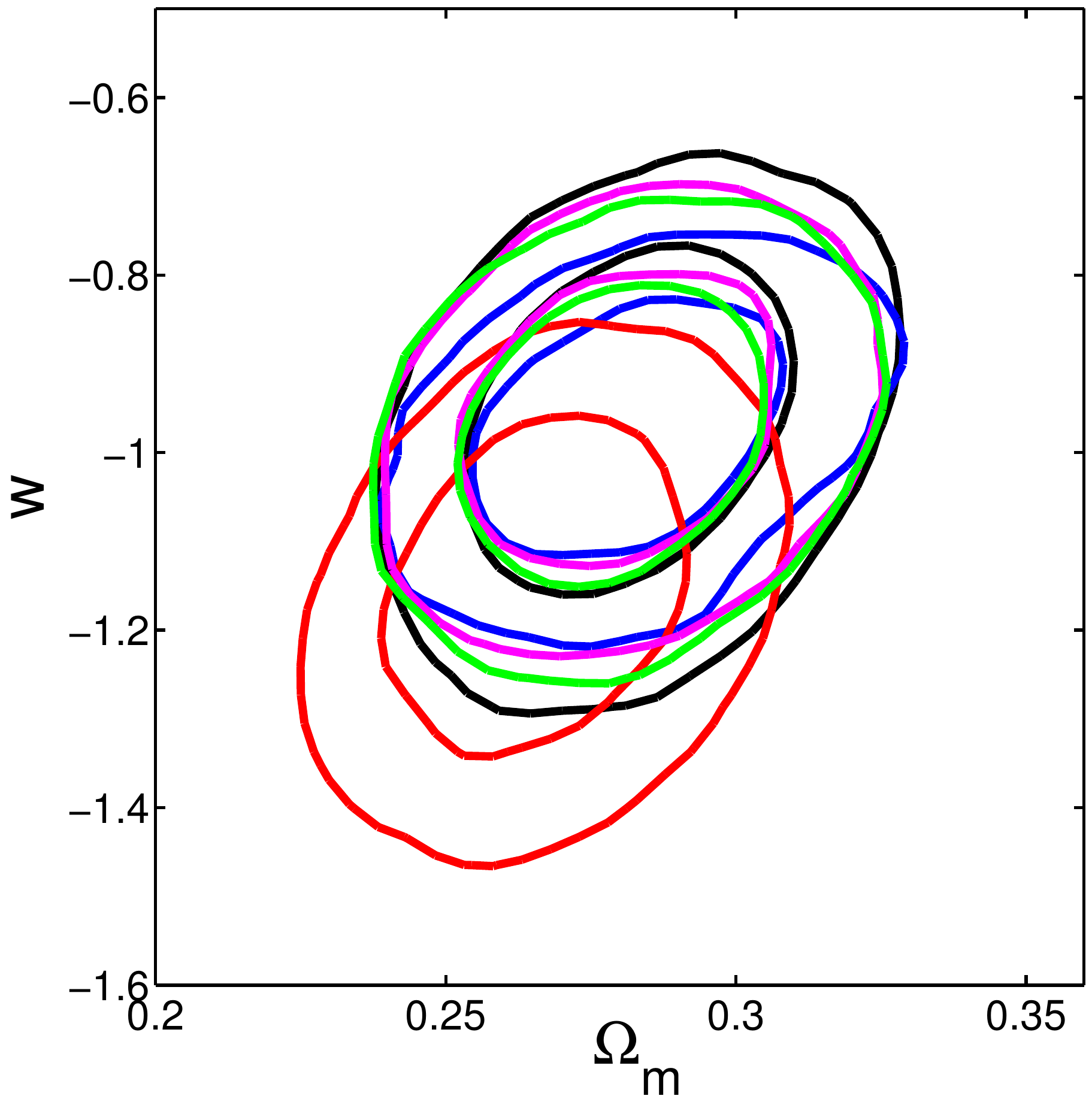}
\caption{\label{cosmology_all}$w$ versus $\Omega_\mathrm{m}$ for the sample of 581 SNe~Ia with host-galaxy measurements, using SNe~Ia+$H_0$+BAO+CMB.  The black contours are uncorrected for SNe~Ia host properties, the blue contours are corrected for the host-galaxy mass step function (with the step at \galmass$=10$). The red, purple and green contours are corrected for the host-galaxy mass, metallicity and age respectively using a linear function, with the slope ($m$) as a free  parameter in the {\sc CosmoMC} fit. The contours enclose 1$\sigma$ and 2$\sigma$ limits on $w$ and $\Omega_\mathrm{m}$. } 
\end{figure*}

Fig.~\ref{cosmology_all} shows the $w$ versus $\Omega_\mathrm{m}$ cosmological contours for all the host-galaxy correction we have investigated in the cosmological analysis, with $H_0$ SH0ES prior \citep{Rie11}, LRGs \citep{Rei10a}, and the full WMAP7 CMB power spectrum \citep{Lar11a} . This clearly shows that all the different fits agree within 1$\sigma$ error contours. The linear correction for the host-galaxy stellar mass, when allowed to vary in the cosmological analysis has the largest effect on the cosmological parameters, shifting to lower values of $\Omega_\mathrm{m}$ and more negative values of $w$.

We also investigate using separate correlations for SNe host-galaxy mass for star-forming or passive galaxies in the cosmological analysis. We find that when using the fixed correlations from Section \ref{mass_met} the derived cosmological parameters are consistent with the results when using a single correction for the host-galaxy mass. Additionally, we tested allowing the slope to vary in the cosmological fit for the passive and star-forming galaxies separately. However, both populations converge to the same value for the slope, and this is consistent with that found when the combined sample was fitted. Thus, we conclude that our current data do not require passive and star-forming galaxies to be separated, but caution that this may become important for the next generation of SN surveys.

\begin{center}      
\begin{sidewaystable*}[!tp] \centering

\vspace{+15cm}
\begin{tabular}{cccccccccccc} 
\hline 
\multicolumn{1}{c}{Correction}  & \multicolumn{4}{c}{Data}  & \multicolumn{2}{c}{Fit parameters} & \multicolumn{5}{c}{Results} \\
\hline
& SNe & $H_0$ & CMB & LRGs  & $m$ &$c$ &$\Omega_{\rm m}$ & $\Omega_{\Lambda}$ & $\Omega_{k}$ & $w$ & Fig.~\ref{cosmology_mass}\\ 
\hline
None & \checkmark & \checkmark  & & & 0 & 0    & 0.363$_{-0.213}^{+0.192}$ &0.637$_{-0.192}^{+0.213}$ & 0 & -1.190$_{-0.699}^{+0.574}$  & Left all \\
\hline
Mass fix  & \checkmark  & \checkmark  & & & -0.078$\pm$0.021  &0.772$\pm$0.211  & 0.330$_{-0.284}^{+0.188}$ & 0.669$_{-0.188}^{+0.284}$ & 0 & -1.154$_{-0.608}^{+0.510}$  & Left top \\
Mass free 1 params & \checkmark  & \checkmark  & &    & -0.123$\pm$0.021 & 0 & 0.287$_{-0.241}^{-0.157}$ & 0.713$_{-0.157}^{+0.241}$ & 0 & -1.316$_{-0.583}^{+0.496}$  & Left top \\
Mass step  & \checkmark  & \checkmark  &  & & log~M~=~10$~{\rm M}_{\rm sun}$& $\Delta M_B$ = 0.091   & 0.394$_{-0.214}^{+0.181}$  & 0.606$_{-0.181}^{+0.214}$ & 0 &  -1.32$_{-0.872}^{+0.728}$& Left top \\
Mass step free  & \checkmark  & \checkmark  & & & log~M~=~9.59$~{\rm M}_{\rm sun}$ & $\Delta M_B$ = 0.262  & 0.389$_{-0.211}^{+0.181}$ & 0.611$_{-0.181}^{+0.211}$ & 0 & -1.321$_{0.803}^{0.659}$  & Left top \\
\hline
Met fix  & \checkmark  & \checkmark  &  & & -0.154$\pm$0.168 &1.320$\pm$1.444 & 0.293$_{-0.247}^{+0.184}$ & 0.707$_{-0.184}^{+0.247}$ & 0 & -1.027$_{-0.434}^{+0.396}$  & Left middle \\
Met free 1 params & \checkmark  & \checkmark  & &  & -0.128$\pm$0.056 & 0 &   0.297$_{-0.250}^{+0.188}$ & 0.703$_{-0.188}^{+0.250}$ & 0 & -1.030$_{-0.446}^{+0.403}$  & Left middle \\
\hline
Age fix  & \checkmark  & \checkmark  & &   &  -0.013$\pm$0.011 & 0.019$\pm$0.022 & 0.312$_{-0.266}^{+0.190}$ &  0.688$_{-0.190}^{+0.266}$ & 0 & -1.089$_{-0.523}^{+0.462}$  &  Left bottom \\
Age free 1 params & \checkmark  & \checkmark  &   & &  -0.025$\pm$0.006 & 0 &  0.316$_{-0.270}^{+0.187}$ & 0.684$_{-0.187}^{+0.270}$ & 0 & -1.106$_{-0.533}^{+0.468}$ &  Left bottom \\
\hline

None & \checkmark  & \checkmark  & \checkmark  & \checkmark    & 0 & 0 & 0.282$_{-0.018}^{+0.018}$ & 0.718$_{-0.020}^{+0.020}$  & -0.0$_{-0.0}^{+0.0}$& -0.970$_{-0.123}^{+0.124}$ & Right top \\
\hline
Mass fix  & \checkmark  & \checkmark  & \checkmark  & \checkmark   & -0.078$\pm$0.021  &0.772$\pm$0.211 &  0.278$_{-0.017}^{+0.018}$ & 0.724$_{-0.019}^{+0.019}$ & -0.001$_{-0.010}^{+0.004}$ & -1.01$_{-0.120}^{+0.120}$  & Right top \\
Mass free 1 params & \checkmark  & \checkmark  & \checkmark  & \checkmark  & -0.126$\pm$0.020 & 0 &  0.266$_{-0.016}^{+0.016}$ & 0.740$_{-0.018}^{+0.018}$ & -0.006$_{-0.007}^{+0.007}$& -1.151$_{-0.121}^{+0.123}$ & Right top \\
Mass step  & \checkmark  & \checkmark  & \checkmark  & \checkmark & log~(M/$~{\rm M}_{\rm sun}$)~=~10 & $\Delta M_B$ = 0.091  & 0.282$_{-0.017}^{+0.017}$ & 0.718$_{-0.017}^{+0.017}$ & -0.001$_{-0.008}^{+0.005}$& -0.975$_{-0.095}^{+0.095}$  & Right top \\
Mass step free  & \checkmark  & \checkmark  & \checkmark  & \checkmark & log~(M$~{\rm M}_{\rm sun}$)~=~9.59 & $\Delta M_B$ = 0.262  & 0.279$_{-0.017}^{+0.017}$ & 0.722$_{-0.019}^{+0.020}$ & -0.001$_{-0.007}^{+0.007}$& -0.986$_{-0.122}^{+0.119}$  & Right: top \\

\hline
Met fix  & \checkmark  & \checkmark  & \checkmark  & \checkmark   &  -0.154$\pm$0.168 &1.320$\pm$1.444 & 0.280$_{-0.017}^{+0.017}$ & 0.729$_{-0.019}^{+0.019}$ & -0.001$_{-0.008}^{+0.007}$& -0.981$_{0-.106}^{+0.105}$  &  Right middle \\
Met free 1 params & \checkmark  & \checkmark  & \checkmark  & \checkmark  &   -0.135$\pm$0.056   & 0  &  0.281$_{-0.017}^{+0.017}$ & 0.719$_{-0.019}^{+0.019}$ & 0.000$_{-0.007}^{+0.007}$ & -0.964$_{-0.107}^{+0.109}$   & Right middle \\

\hline
Age fix  & \checkmark  & \checkmark  & \checkmark  & \checkmark   & -0.013$\pm$0.011 & 0.019$\pm$0.022 & 0.281$_{-0.017}^{+0.017}$ & 0.719$_{-0.018}^{+0.018}$ & 0.001$_{-0.007}^{+0.006}$ & -0.975$_{-0.103}^{+0.109}$  &  Right bottom \\
Age free 1 params & \checkmark  & \checkmark  & \checkmark  & \checkmark  &  -0.024$\pm$0.006 & 0 &   0.280$_{-0.017}^{+0.017}$ & 0.721$_{-0.018}^{+0.019}$ & -0.001$_{-0.007}^{+0.007}$ & -0.979$_{-0.110}^{+0.108}$ &  Right bottom \\

\hline 
\end{tabular}
\caption{\label{cosmo_table}  Summary of the cosmological fits presented in Section~\ref{sect:cosmo}.}
\end{sidewaystable*}
\end{center}

\section{Conclusions}
\label{sect:discussion}

In this paper, we have investigated correlations between SNe~Ia light curves and their host galaxies and look at the effect on the cosmological constraints. For this we have used the sample of 581 photometrically-classified SNe~Ia from \cite{Cam13}. This sample was assembled from three years of photometry from the SDSS-II SN Survey, together with BOSS spectroscopy of the host galaxies of transients. We use the stellar population parameters derived from the BOSS DR10 results \citep{Ahn12} and calculate the metallicities from the measured line fluxes. The stellar masses are derived using SED fitting with the \citet{Mar13} models. Compared to previous studies, our sample is larger by a factor of $>4$. We also cover a wider range of redshifts (up to $z\sim0.5$), which is directly applicable to cosmological measurements.

Our main results are as following.
\begin{itemize}

\item We measure a significant correlation ($>5\sigma$) between the host-galaxy stellar-mass and the SNe~Ia HR, consistent with previous studies. The correlation between host-galaxy stellar mass and the SNe~Ia HR is likely to be due to an underlying physical property, which the mass (which is easier to measure) is simply acting as a proxy for. We find that whether the host mass is fit with a linear fit or a step function will alter the derived cosmological parameters.

\item We find a weak correlation ($1.4\sigma$) between the host-galaxy metallicity and the SNe~Ia HR. Comparing the slope of this correlation to previous studies, we find that it is consistent, and that all studies show the same direction of trend.

\item We consider a small sample of host-galaxy spectra taken at the position of the SN, but unfortunately the sample size is too small to draw meaningful conclusions.

\item We find that the slope of the correlation between HR and host-galaxy mass is {\it steeper} for low-metallicity galaxies.

\item We test the effects of either applying a fixed linear correction (based on either host-galaxy stellar mass, metallicity or age) to the distance modulus, or allowing an additional free parameter to account for this within a cosmological fit. We see a shift towards a more negative equation-of-state parameter $w$ and a shift to lower values of $\Omega_\mathrm{m}$ when including a correction for the host-galaxy stellar mass, especially when the relation coefficients are fit simultaneous with the cosmological parameters. The shift with host-galaxy stellar-mass correction is consistent with previous studies \citep{Sul06}, and we also see a small reduction in the size of the cosmological contours. When a fixed correction for the host-galaxy stellar mass is applied it biases the $\alpha$ and $\beta$ parameters.

\item We find that the host-galaxy stellar mass has a much more significant effect on the cosmological parameters than the host-galaxy metallicity or age.
\end{itemize}

As current and next generation surveys move towards a goal of 1 per cent cosmology, small systematic effects such as the host-galaxy mass and metallicity correlations considered here become ever more important. While it is possible to measure these systematic effects and correct for them, to properly account for the covariances and degeneracies between SNe~Ia and host-galaxy parameters it is better to fit and solve for these effects simultaneously with cosmological parameters.

We also suggest that a future avenue for this work could be the inclusion of host morphologies from Galaxy Zoo. 17 of our host galaxies were included in the GZ1 or GZ2 catalogues from Galaxy Zoo \citep{Lin08,Wil:13}\footnote{Publicly available at http://data.galaxyzoo.org/}. This sample is only a small subset of our full sample as most of our hosts are unresolved with $r\gtrsim~18$ mag. However, this would be extremely interesting to investigate in the future with a lower redshift sample.

\section{Acknowledgements}

We are indebted to the anonymous referee, who made many constructive and helpful suggestions which have improved this work.
We thank Bethan James, Bob Nichol, Rubina Kotak, Max Pettini, Mark Sullivan, Clare Worley, Sergey Koposov and Andy Casey for useful discussion and advice. We thank Rachel Wolf, Chris D'Andrea, Ravi Gupta, Masao Sako, Roberto Trotta and Hikmatali Shariff for helpful and constructive discussions and comments on this work.

This work was partly supported by the European Union FP7 programme through ERC grant number 320360.

Funding for SDSS-III has been provided by the Alfred P. Sloan Foundation, the Participating Institutions, the National Science Foundation, and the U.S. Department of Energy Office of Science. The SDSS-III web site is http://www.sdss3.org/.

SDSS-III is managed by the Astrophysical Research Consortium for the Participating Institutions of the SDSS-III Collaboration including the University of Arizona, the Brazilian Participation Group, Brookhaven National Laboratory, Carnegie Mellon University, University of Florida, the French Participation Group, the German Participation Group, Harvard University, the Instituto de Astrofisica de Canarias, the Michigan State/Notre Dame/JINA Participation Group, Johns Hopkins University, Lawrence Berkeley National Laboratory, Max Planck Institute for Astrophysics, Max Planck Institute for Extraterrestrial Physics, New Mexico State University, New York University, Ohio State University, Pennsylvania State University, University of Portsmouth, Princeton University, the Spanish Participation Group, University of Tokyo, University of Utah, Vanderbilt University, University of Virginia, University of Washington, and Yale University.

\appendix

\section{Host galaxy correlations without upper limits}
\label{host_gal_no_lims}
Our sample of SN host galaxies has 332 with measured metallicities and 259 with upper limits. \citet{Kel07} investigated using upper limits or `censored' data in the dependent variable but suggested it was easiest to remove and refit the data when the limits were in the independent variable. We follow this procedure and remove the upper limits for our metallicity (or SFR) sample and refit the correlations. The results from these fits can be see in Table~\ref{host_fits_2_line_no_limits} and are shown as the cyan dashed line on Fig.~\ref{correlations}. These are all consistent with the analysis carried out including the upper limits. However, some of the correlations slightly change their significance. Most notably the colour versus sSFR correlations becomes more significant (1.45$\sigma$  to 4$\sigma$) when the upper limits are excluded.

\begin{table*}
\hspace{-1cm}
\centering
\begin{tabular}{cccccc}
 \hline
$x$ & $y$ & $m$ & $c$ &sig & \% \\
\hline 
Metallicity  & HR corr       & $-0.205\pm0.182$~mag/dex & 1.782$\pm$1.573~mag  & 1.5$\sigma$ &  87.18\% \\ 
-       & HR uncorr  &  $-0.205\pm0.182$~mag/dex & 1.782$\pm$1.573~mag  & 1.5$\sigma$&  87.18\% \\
-       & $x_1$               &  $-0.830\pm0.539$~mag/dex & 7.296$\pm$4.636~mag  & 1.8$\sigma$ &  93.81\% \\
-       & colour         &  $0.023\pm0.055$~mag/dex  & -0.214$\pm$0.474~mag   & 0.9$\sigma$ &  66.67\% \\
\hline
SFR & HR corr & $0.051\pm0.059$~mag/$\mathrm{log}({\rm M}_\odot{\rm yr}^{-1})$ & -0.014$\pm$0.018~mag &  $1.4\sigma$   & 81.43\% \\
-  & HR uncorr & $0.053\pm0.061$~mag/$\mathrm{log}({\rm M}_\odot{\rm yr}^{-1})$ & -0.045$\pm$0.018~mag &  $1.3\sigma$   & 79.52\% \\
-     & $x_1$ & $1.286\pm0.183$~mag/$\mathrm{log}({\rm M}_\odot{\rm yr}^{-1})$  & -0.214$\pm$0.0554~mag &  $>5\sigma$   & 100\% \\
-    & colour & $0.034\pm0.019$~mag/$\mathrm{log}({\rm M}_\odot{\rm yr}^{-1})$  & -0.024$\pm$0.006~mag &  $2\sigma$   & 95.51\% \\
       \hline
sSFR  & HR corr & $0.081\pm0.371$~mag/$\mathrm{log}({\rm yr}^{-1})$ & 0.852$\pm$3.917~mag  &  $0.8\sigma$   & 58.82\% \\
-  & HR uncorr & $0.063\pm0.327$~mag/$\mathrm{log}({\rm yr}^{-1})$  & 0.630$\pm$3.479~mag  &  $0.8\sigma$   & 56.66\% \\
-       & $x_1$ & $2.805\pm0.435$~mag/$\mathrm{log}({\rm yr}^{-1})$    & 29.794$\pm$4.578~mag &  $>5\sigma$   & 100\% \\
-    & colour & $0.655\pm0.211$~mag/$\mathrm{log}({\rm yr}^{-1})$   & 6.941$\pm$2.241~mag &  $4\sigma$   & 99.995\% \\

  \hline    
\end{tabular}
 \caption{\label{host_fits_2_line_no_limits} Summary of the fits and significance of the correlations between the host-galaxy properties and the SNe~Ia parameters with AGN removed by the `two-line' diagnostic and only SNe~Ia with measured host parameters considered in the fits (upper limits excluded). $m$ is the slope of the correlation, $c$ is the intercept with the $y$-axis. The columns `sig' and `\%' show the significance of the correlation, both in units of $\sigma$ and in the percentage of samples from the posterior distribution of slopes which lie above or below zero.}
 
\end{table*}

\section{Cosmological parameter correlations}
\label{appendix_b}

\begin{figure*}
\includegraphics[width=1.0\linewidth,height=0.4\linewidth]{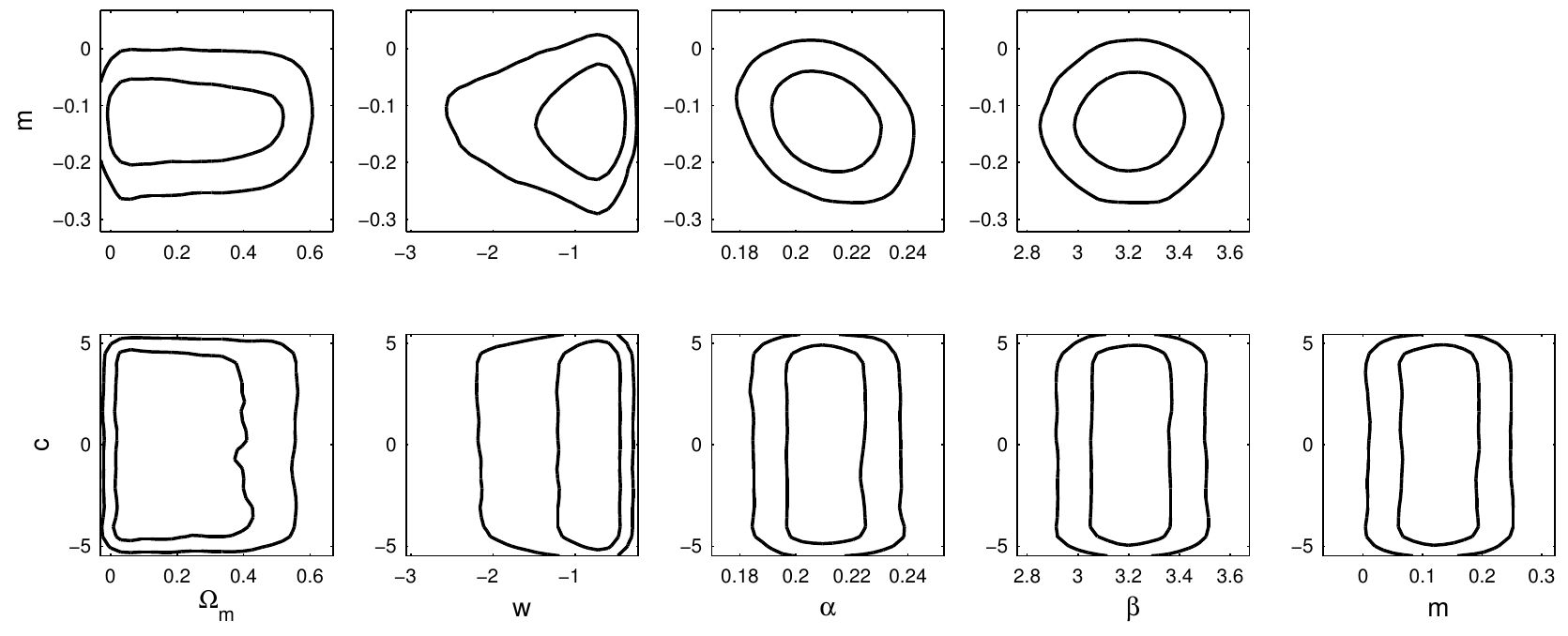}
\caption{\label{m_c_w_mass} The slope ($m$) and intercept ($c$) from the linear correlation with host-galaxy stellar mass  ($m \times M_\mathrm{host}+c$) when they are allowed to vary as free parameters in the {\sc CosmoMC} fit, versus the main cosmological and SNe~Ia parameters. In each panel the contour lines enclose 1$\sigma$ and 2$\sigma$ of the posterior distribution. As can be seen, the intercept $c$ is unconstrained by the data.}
\end{figure*}

In Section~\ref{sect:cosmo}, we investigate using the correlations with host-galaxy stellar mass as an additional parameter in the cosmological analysis. 

Fig. ~\ref{m_c_w_mass} shows the correlations between the slope ($m$) and intercept ($c$) of the correction for the host-galaxy mass correlation with the cosmological parameters when the slope and offset are allowed to vary in the {\sc CosmoMC} fit, for the sample of 581 SNe~Ia with measured host-galaxy stellar mass, using only SNe~Ia data plus the prior on $H_0$. We see that the offset is not constrained, suggesting that having this extra degree of freedom is not required by the current data. We hence rerun our cosmological analysis allowing only one free parameter to account for the host-galaxy mass.

\begin{figure*}
\includegraphics[width=1.0\linewidth,height=1.0\linewidth]{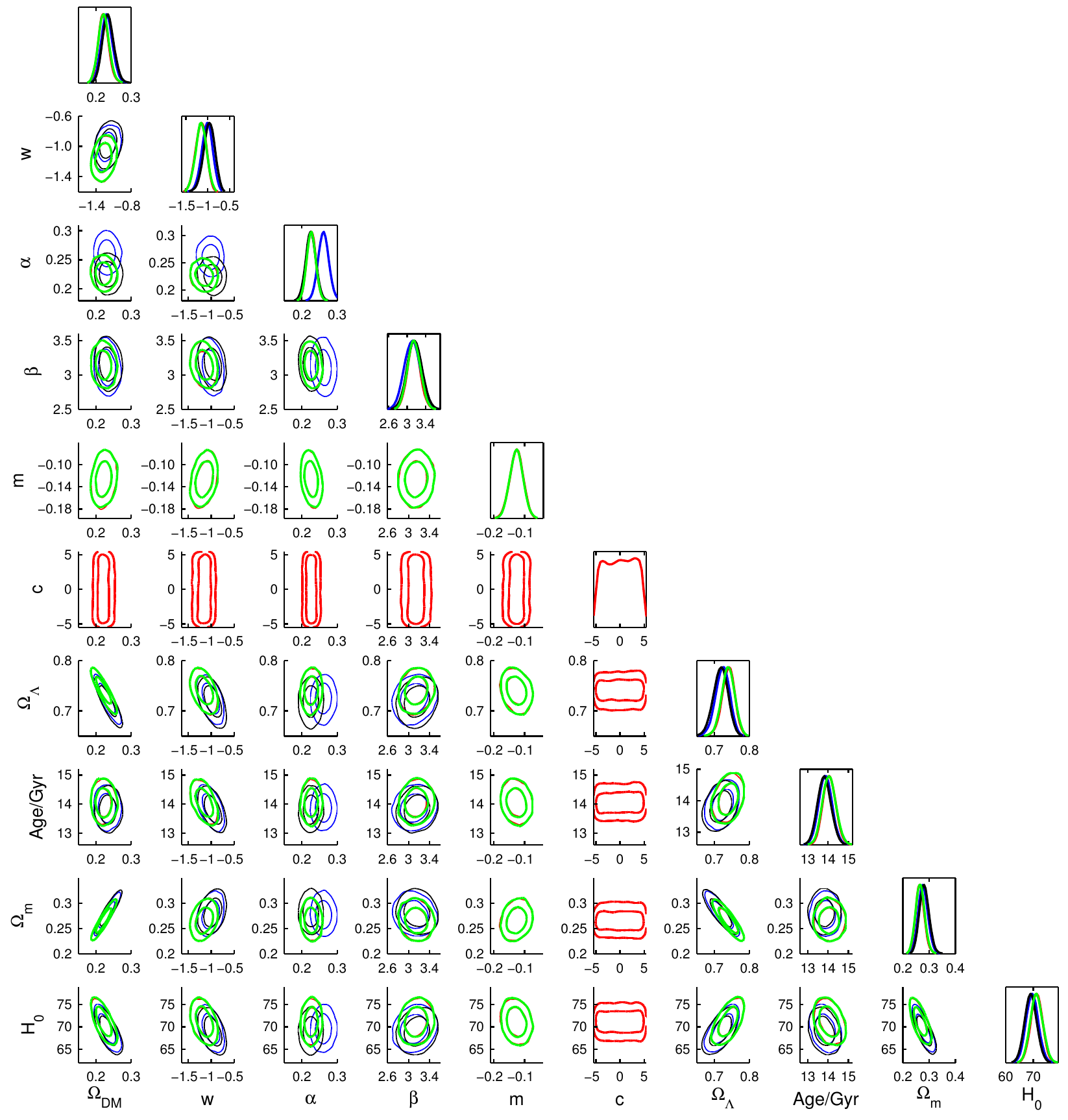}
\caption{\label{cosmo_all_mass_all} All potential correlations between parameters in the {\sc CosmoMC} fit, for the sample of 581 SNe~Ia with measured host-galaxy stellar mass, using SNe~Ia~+~$H_0$~+~BAO~+~CMB. The black contours are uncorrected. The blue contours are corrected for the host-galaxy stellar mass using the best-fitting linear function, with $m=-0.078$ and $c=0.772$. The red contours have $m$ and $c$ as free parameters in the {\sc CosmoMC} fit, while the green contours have only $m$ as a free parameter. The red contours are nearly always obscured by the green contours, showing that the additional offset parameter is not needed. Contours  enclose 1$\sigma$ and 2$\sigma$ of the posterior distribution.}
\end{figure*}

Fig. ~\ref{cosmo_all_mass_all} shows all the potential correlations between recovered parameters in the {\sc CosmoMC} fit, for the sample of 581 SNe~Ia with measured host-galaxy stellar mass, using data from SNe~Ia~+~$H_0$~+~BAO~+~CMB. It is evident that when a fixed correction for the mass correlation is used, the best-fitting $\alpha$ (and to a lesser extent, $\beta$) recovered from the cosmological analysis is shifted to higher values. The plot of $m$ versus $\alpha$ suggests that these two parameters are degenerate. This might be expected as SNe~Ia in more massive galaxies tend to be broader (and hence have higher $x_1$). This suggests that deriving a correlation for the host-galaxy mass with all the cosmological parameters fixed, and then applying this may create a bias in the analysis, which is somewhat compensated for by the change in $\alpha$. As an alternative, we suggest that it is safer to allow the correction for the host-galaxy mass to be an additional free parameter which is solved for simultaneously in the cosmological fit, rather than measuring it independently and attempting to `correct' the data. In Fig. ~\ref{cosmo_all_mass_all} the fits with both the slope and intercept of the mass correlation as free parameters, and those where only the slope was allowed to vary, are almost indistinguishable. This strengthens our conclusion that the additional parameter for the intercept $c$ is not required. The fits with the correction for the mass as a free parameter are also found to shrink the 1$\sigma$ and 2$\sigma$ error contours for many of the derived cosmological parameters and in some cases changes the best-fitting value.

\begin{figure*}
\includegraphics[width=1.0\linewidth,height=1.0\linewidth]{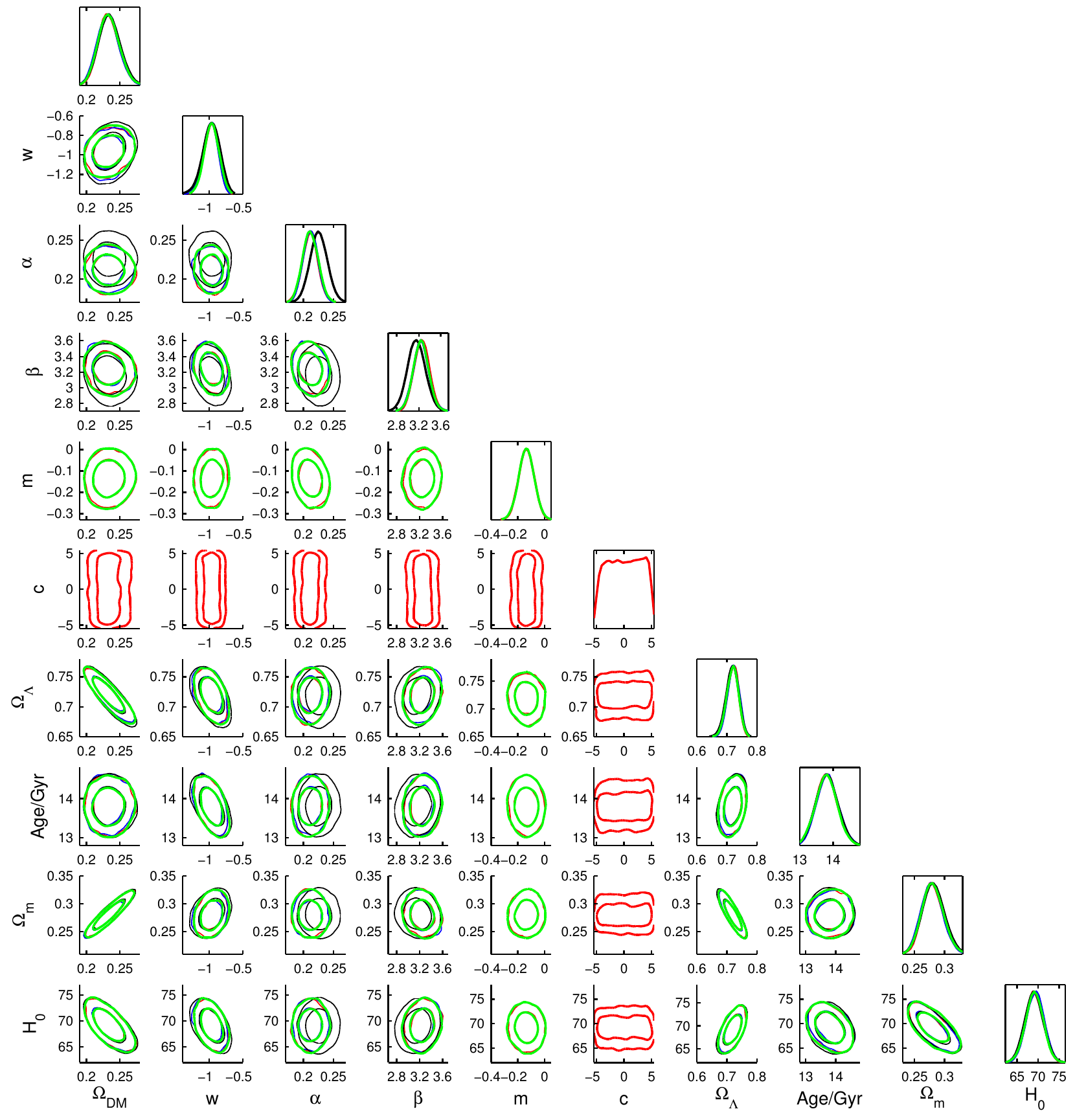}
\caption{\label{cosmo_all_met_others}All the potential correlations between parameters in the {\sc CosmoMC} fit, for the sample of 581 (259 are limits from the continuum flux at the position of the spectral emission lines) SNe~Ia with measured host-galaxy metallicity, using only SNe~Ia~+~$H_0$~+~BAO~+~CMB. The black contours are uncorrected, the blue contours are corrected for the host-galaxy metallicity using the best-fitting linear function ($m=-0.154$ and $c=1.320$), the red contours have $m$ and $c$ as free parameters in the {\sc CosmoMC} fit, the green contours have only $m$ as a free parameter. The red contours are nearly always obscured by the green contours, showing that the additional offset parameter is not needed. }
\end{figure*}

The cosmological analysis was repeated using a correction for host-galaxy metallicity. Fig.~\ref{cosmo_all_met_others} shows all the possible correlations between the {\sc CosmoMC} fitted parameters, including CMB and BAO constraints as well as SNe~Ia and a prior on $H_0$. Here, we see the same small change in $\alpha$ when the free parameter for metallicity is included in the fit, however, the change in $\beta$ is now less marked than when using the mass correction. The size of the error contours and their location is similar for all the other cosmological fits. Adding a correction for metallicity either as a fixed value or as free parameter does not seem to improve the cosmological analysis when combined with other cosmological probes.

%%-----------Appendix E
\section{SNe~Ia and host-galaxy data}
\label{Appendix_data}

In Table~\ref{DATA_table} we present the key information used in this paper for our sample of \nofinal\ photometrically--classified SNe~Ia \citep{Cam13} with host galaxy information, the table can be electronically downloaded from http://www.mnras.oxfordjournals.org/lookup/suppl/doi:10.1093/mnras/stw115/-/DC1. Table \ref{DATA_table} lists the columns within this table. 
\begin{table}
\hspace{-0.8cm}
\begin{tabular}{lc|c|c|}
  \hline
Field &  Property \\
  \hline
1	& SN ID\\
2	& Host-galaxy redshift\\
3	& Host-galaxy redshift error\\
4	&	SN RA [Deg J2000]\\
5	&	SN Dec [Deg J2000]	\\
6  & SN SALT2 X0  [mag]\\
7  & SN SALT2 X0 error  [mag]\\
8  & SN SALT2 X1  [mag]\\
9    & SN SALT2 X1 error  [mag]\\
 10     & SN SALT2 colour  [mag]\\
11     & SN SALT2 colour error [mag]\\ 
 12    & SN SALT2 X0 X0 covariance \\ 
  13   & SN SALT2 X0 X1 covariance \\ 
    14   & SN SALT2 X0 colour covariance \\ 
  15       & SN SALT2 X1 X1 covariance \\ 
   16          & SN SALT2 X1 colour covariance\\ 
    17        & SN SALT2 color colour covariance  \\ 
18       & SN $\mu$ uncorrected [mag]\\     
 19   	 & SN $\mu$ corrected [mag]\\
20  & SN $\mu$ error [mag]\\  
21  	& SN best-fitting cosmology estimated $\mu$   [mag]\\
22  & SN best-fitting cosmology estimated $\mu$ error  [mag]\\       
23  	& SN HR from $\mu$ uncorrected [mag]\\  
24  	 & SN HR from $\mu$ corrected [mag]\\
25 &  HR error [mag]\\
26	& Host-galaxy ObjID (DR8)\\
27	 & Host-galaxy type \\
28	 & Host-galaxy metallicity [mag/dex]\\	
29	 & Host-galaxy metallicity error [mag/dex]\\
30	 & Host-galaxy metallicity type (measured or upper limit) \\
31	 & Host-galaxy mass [\magpergalmass]\\
32	 & Host-galaxy mass error [\magpergalmass]\\
33	 & Host-galaxy SFR [mag/$\mathrm{log}({\rm M}_\odot{\rm yr}^{-1})$]\\
34	 & Host-galaxy SFR error [mag/$\mathrm{log}({\rm M}_\odot{\rm yr}^{-1})$]\\
35	 & Host-galaxy SFR type (measured or upper limit) \\
36	 & Host-galaxy sSFR [mag/$\mathrm{log}({\rm yr}^{-1})$]\\
37	 & Host-galaxy sSFR error [mag/$\mathrm{log}({\rm yr}^{-1})$]\\
38	 & Host-galaxy age [mag/$\mathrm{log}$(Gyr)]\\
39	 & Host-alaxy age error [mag/$\mathrm{log}$(Gyr)]\\
\hline
\end{tabular}
\caption{Table illustrating the data available for the 581 SNe~Ia with host galaxy data presented in this paper in our electronic table at http://www.mnras.oxfordjournals.org/lookup/suppl/doi:10.1093/mnras/stw115/-/DC1. The 15 SNe with entries of * in the electronic table are cases where there is no photometric object ID for the host galaxy in DR8; these galaxies do appear in the co-added images, and hence within this catalogue we quote the HostID.}

\label{DATA_table}
\end{table}

\bibliographystyle{plainnat}

\begin{thebibliography}{144}
\expandafter\ifx\csname natexlab\endcsname\relax\def\natexlab#1{#1}\fi



\bibitem[Ahn et al.(2012)]{Ahn12}
	Ahn C.P. et al., 2012, \apjs, 203, 21
	
\bibitem[Baldwin et al.(1981)]{Bal81}
	Baldwin J.A., Phillips M.M., Terlevich R.,1981, \pasp, 93, 5

\bibitem[Betoule et al. (2014)]{Bet14}
	Betoule M. et al., 2014, \aaps, 568, A22

\bibitem[Brinchmann et al. (2004)]{Bri04}
	Brinchmann J., Charlot S., White S.D.M., Tremonti C., Kaumann G., Heckman T., Brinkmann J., 2004, \mnras, 351, 1151

\bibitem[Bolzonella et al.(2000)]{Bol00}
	Bolzonella M., Miralles J.-M., Pell{\'o} R., 2000, \aaps, 363, 476

 \bibitem[Calzetti (2001)]{Cal01}
	Calzetti D., 2001, \pasp, 113, 1449

\bibitem[Carter et al. (2001)]{Car01}
	Carter B.J., Fabricant D.G., Geller M.J., Kurtz M.J., McLean B,. 2001, \apj, 559, 606 

\bibitem[Campbell et al. (2013)]{Cam13}
	Campbell H. et al., 2013, \apj, 763, 88 

\bibitem[Childress et al. (2013a)]{Chi13a}
	Childress M. et al.\ 2013a, \apj, 770, 107
	
\bibitem[Childress et al. (2013b)]{Chi13b}
	Childress M. et al.\ 2013b, \apj, 770, 108 

\bibitem[Conroy et al. (2009)]{Con09} 
	Conroy C., Gunn J. E., White M., 2009, ApJ, 699, 486

\bibitem[Conroy \& Gunn (2010)]{Con10}
	Conroy C.,  Gunn J. E., 2010, ApJ, 712, 833

\bibitem[D'Andrea et al. (2011)]{DAn11} 
	D'Andrea C.B. et al., 2011, \apj, 743, 172

\bibitem[Dawson et al. (2013)]{Daw13}
	Dawson K.S. et al., 2013 \aj, 145, 10

\bibitem[Dilday et al. (2008)]{Dil08}
	Dilday B. et al., 2008, \apj, 682, 262

\bibitem[Dilday et al. (2010)]{Dil10}
	Dilday B. et al., 2010, \apj, 715, 1021
 
 \bibitem[Fioc \& Rocca-Volmerange(1997)]{Fio97}
	Fioc M., Rocca-Volmerange B., 1997 , A\&A, 326, 950

 \bibitem[Fioc \& Rocca-Volmerange(1999)]{Fio99}
	Fioc M., Rocca-Volmerange B., 1999, (astro-ph/9912179)
 
\bibitem[Foley \& Kasen (2011)]{Fol11}
	Foley R.J., Kasen D., 2011, \apj, 729, 55 

 \bibitem[Foley (2012)]{Fol12}
	Foley R.J., 2012, \apj, 748, 127

\bibitem[Frieman et al. (2008)]{Fri08}
	Frieman J.A. et al., 2008, \aj, 135, 338

\bibitem[Galbany et al. (2012)]{Gal12}
	Galbany L. et al., 2012 \apj, 755, 125
	
\bibitem[Gunn et al. (1998)]{Gun98}	
	Gunn J.E et al., 1998, \aj, 116, 3040

\bibitem[Gupta et al. (2011)]{Gup11}
	Gupta R.R. et al., 2011 \apj, 740, 92

 \bibitem[Guy et al. (2007)]{Guy07}
 	Guy J. et al., 2007 \aaps, 466,11
 
 \bibitem[Guy et al. (2010)]{Guy10} 
 	Guy J. et al., 2010, \aaps, 523, A7
	
\bibitem[Hayden et al. (2010)]{Hay10}
	Hayden B.T. et al., 2010 \apj, 722, 1691

\bibitem[Hayden et al. (2013)]{Hay13}
	Hayden B.T., Gupta R.R., Garnavich P.M., Mannucci F., Nichol R.C, Sako M., 2013 \apj, 764, 191
	
	\bibitem[Hicken et al. (2009)]{Hic09}
	Hicken M., Wood-Vasey W. M., Blondin S., Challis P., Jha S., Kelly P. L., Rest A., Kirshner R., 2009, ApJ, 700, 1097
		
\bibitem[Hillebrandt \& Niemeyer(2000)]{Hil00}
	Hillebrandt W., Niemeyer J.C., 2000, \araa, 38, 191 
	

\bibitem[Howell et al (2009)]{How09}
	Howell D.A. et al., 2009 \apj, 691, 661
	
\bibitem[Johansson et al. (2013)]{Joh13} 
	Johansson J, et al., 2013 \mnras,  435,1680

\bibitem[Jones et al. (2015)]{Jon15}
 Jones D.O., Riess A.G, Scolnic D.M., 2015, AAS Meeting, 227, 139.08
 
\bibitem[Kelly(2007)]{Kel07}
	Kelly B.C., 2007, \apj, 665, 1489 

\bibitem[Kelly et al. (2010)]{Kel10}
	Kelly P.L., Hicken M., Burke D.L., Mandel K.S., Kirshner R.P., 2010, ApJ, 715, 743
	
 \bibitem[Kelly et al. (2015)]{Kel15}
	Kelly P.L., Filippenko A.V., Burke D. L., Hicken M., Ganeshalingam M., Zheng W., 2015, Science, 347, 1459	
	
\bibitem[Kennicutt (1998)]{Ken98}
	Kennicutt R. C.,1998, ARA\&A, 36, 189
	
\bibitem[Kessler et al. (2009)]{Kes09}
	Kessler R. et al., 2009, \apjs, 185, 32

\bibitem[Kewley et al.(2001)]{Kew01}
	Kewley L.J., Dopita M.A., Sutherland R.S., Heisler C.A., Trevena J., 2001, \apj, 556, 121
    
%\bibitem[Kewley et al.(2002)]{Kew02}
     
\bibitem[Kim et al. (2014)]{Kim14}
	Kim A.G. et al., 2014, \apj, 784, 51
  
\bibitem[Konishi et al. (2011)]{Kon11}
	Konishi K. et al., 2011, (arXiv:1101.1565)

\bibitem[Kudritzki et al. (2014)]{Kud14}
	Kudritzki R., Urbaneja M.A., Bresolin F., Hosek M.W., Przybilla N., 2014, \apj, 788, 56

\bibitem[Lampeitl et al. (2010a)]{Lam10a}
	Lampeitl H. et al., 2010a, \mnras, 401, 2331

 \bibitem[Lampeitl et al. (2010b)]{Lam10b} 
	Lampeitl H. et al., 2010b, \apj, 722, 566

\bibitem[Larson et al. (2011)]{Lar11a}
	Larson D. et al., 2011, \apjs, 192, 16

\bibitem[Lewis \& Bridle (2002)]{Lew02}
	Lewis A., Bridle S., 2002, \prd, 66, 103511

\bibitem[Li \& White (2009)]{Li09}
	Li C., White S.D.M., 2009, \mnras, 398, 2177

\bibitem[Li et al.(2011)]{Li11}
	Li W. et al.\ 2011, \mnras, 412, 1441 

 \bibitem[Liang et al. (2007)]{Lia07}
	Liang Y. C., Hammer F., Yin S. Y., Flores H., Rodrigues M., Yang Y. B. 2007, A\&A, 473, 411

\bibitem[Lintott et al. (2008)]{Lin08}
	Lintott C.J. et al., 2008, \mnras, 389, 1179

\bibitem[Maguire et al.(2013)]{Mag13}
	Maguire K., et al.\ 2013, \mnras, 436, 222 

\bibitem[Maraston et al. (2006)]{Mar06}	
	Maraston C., Daddi E., Renzini A., Cimatti A., Dickinson M., Papovich C., Pasquali A., Pirzkal N., 2006, \apj, 652, 85

\bibitem[Maraston et al. (2013)]{Mar13}
	Maraston C. et al., 2013, \mnras, 435, 2764

\bibitem[Miller et al. (2003)]{Mil03}
	Miller C.J., Nichol R.C., Gomez P.L., Hopkins A.M., Bernardi M., 2003, \apj,  597, 142

\bibitem[Neill et al. (2009)]{Nei09}
	Neill J.D. et al., 2009, \apj, 707, 1449

 \bibitem[Nordin et al. (2011)]{Nor11}
	Nordin J. et al., 2011, \apj, 734, 42
	
\bibitem[Ostman et al. (2011)]{Ost11}
	\'Oestman L. et al., 2014, \aaps, 526, A28
	
\bibitem[Olmstead et al. (2014)]{Olm14}
	Olmstead M.D. et al. 2014, \aj, 147, 75
	
\bibitem[Pan et al.(2014)]{Pan14}
	Pan Y.-C. et al., 2014, \mnras, 438, 1391 

\bibitem[Perlmutter et al.(1999)]{Per99}
	Perlmutter S. et al., 1999, \apj, 517, 565 

\bibitem[Pettini \& Pagel(2004)]{Pet04}
	Pettini M., Pagel B.E.J., 2004, \mnras, 348, L59 
	
\bibitem[Phillips(1993)]{Phi93}
	Phillips M.M., 1993, \apjl, 413, L105 

\bibitem[Pilyugin(2001)]{Pil01}
Pilyugin, L. S. 2001, A\&A, 374, 412

\bibitem[Pilyugin \& Thuan (2005)]{Pil05}
Pilyugin L. S., Thuan T. X. 2005, ApJ, 631, 231


\bibitem[Planck Collaboration XIII (2015)]{Pla15}
	Planck Collaboration XIII \ 2015, (arXiv:1502.01589) 
	
\bibitem[Reid et~al.(2010)]{Rei10a}
	Reid B.A. et al., 2010, \mnras, 404, 60

\bibitem[Riess, Press \& Kirshner(1996)]{Rie96}
	Riess A.G., Press W.H., Kirshner R.P., 1996, \apj, 473, 88

\bibitem[Riess et al.(1998)]{Rie98}
	Riess A.G. et al., 1998, \aj, 116, 1009 

\bibitem[Riess et~al.(2011)]{Rie11}
	Riess A.G. et al., 2011, \apj, 730, 119

\bibitem[Rigault et al. (2013)]{Rig13}
	Rigault M. et al., 2013, \aaps, 560, 66

\bibitem[Rigault et al. (2015)]{Rig15}
	Rigault M. et al., 2015, ApJ, 802, 20

\bibitem[Sako et al. (2008)]{Sak08}
	Sako M. et al., 2008, \aj, 135, 348
		
\bibitem[Sako et al.(2011)]{Sak11}
	Sako M. et al., 2011, \apj, 738, 162
		
\bibitem[Sako et al.(2014)]{Sak14}
	Sako M. et al., 2014, arXiv:1401.3317 

\bibitem[Sarzi et al.(2006)]{Sar06}
	Sarzi M. et al., 2006, \mnras, 366, 1151

\bibitem[Schmidt et al.(1998)]{Sch98}
	Schmidt B.P. et al., 1998, \apj, 507, 46 

\bibitem[Smith et al. (2012)]{Smi12}
	Smith M. et al., 2012, \apj, 755, 61

\bibitem[Sollerman et al. (2009)]{Sol09}
	Sollerman J. et al., 2009, \apj, 703, 1374
 
 \bibitem[Sullivan et al. (2006)]{Sul06}
	Sullivan M. et al., 2006, \aj, 131, 960

\bibitem[Sullivan et al.(2010)]{Sul10}
	Sullivan M. et al., 2010, \mnras, 406, 782 

\bibitem[Sullivan et al.(2011)]{Sul11}
	Sullivan M. et al., 2011, \apj, 737, 102 
	
\bibitem[Thomas et al.(2013)]{Tho13}
	Thomas D. et al., 2013, \mnras, 431, 1383 
		
	
		
\bibitem[Willett et al. (2013)]{Wil:13}
	Willett K.W. et al., 2013 \mnras, 435, 2835
	
\bibitem[Wolf et al. (2015)]{Wol15}
	Wolf C. et. al., 2015, submitted 
		
	\bibitem[Yin et al. (2007)]{Yin07}
	Yin S. Y., Liang Y. C., Zhang B. 2007, ASP Conf, 373, 686
	
\bibitem[Zheng et al. (2008)]{Zhe08}
	Zheng C. et al., 2008, \aj, 135, 1766



\end{thebibliography}

\bsp

%\clearpage

\end{document}